\documentclass[a4paper,12pt]{article}
\usepackage{amsmath}
\usepackage{amsbsy}
\usepackage{amsthm}
\usepackage{amsfonts}
\usepackage{amssymb}
\usepackage{graphicx}
\usepackage{epstopdf}
\usepackage{tikz}

\newtheorem{theorem}{Theorem}
\newtheorem{lemma}{Lemma}

\newtheorem{innercustomthm}{Theorem}
\newenvironment{customthm}[1]
  {\renewcommand\theinnercustomthm{#1}\innercustomthm}
  {\endinnercustomthm}
  
\newtheorem{innercustomthml}{Lemma}
\newenvironment{customthml}[1]
  {\renewcommand\theinnercustomthml{#1}\innercustomthml}
  {\endinnercustomthml}
  
\numberwithin{equation}{section}

\let\a=\alpha \let\b=\beta         \let\d=\delta     \let\e=\varepsilon
             
\let\m=\mu    \let\n=\nu              \let\p=\pi        
\let\s=\sigma \let\t=\tau            \let\c=\chi
        
\let\G=\Gamma \let\D=\Delta       \let\L=\Lambda

\def\ss{{\underline\s}}

\def\DD{{\cal D}}\def\PP{{\cal P}}

\def\ZZZ{{\mathbb Z}}\def\NNN{{\mathbb N}}

\let\dpr=\partial

\let\==\equiv

\def\media#1{{\left\langle#1\right\rangle}}

\def\V#1{{\bf #1}}
\def\Vn{{\bf n}}
\def\xx{{\bf x}}
\def\yy{{\bf y}}

\def\be{\begin{equation}}
\def\ee{\end{equation}}
\def\bea{\begin{eqnarray}}\def\eea{\end{eqnarray}}

\pgfdeclarelayer{background}
\pgfsetlayers{background,main}

\begin{document}

\title{Periodic striped ground states in Ising models with competing interactions}

\author{\vspace{5pt} Alessandro Giuliani$^{1}$ and Robert
  Seiringer$^{2}$\\
  \vspace{-4pt}\small{$^{1}$Dipartimento di Matematica Universit\`a di Roma Tre} \\ \small{
L.go S. L. Murialdo 1, 00146 Roma, Italy}\\
  \vspace{-4pt}\small{$^{2}$Institute of Science and Technology Austria,}\\
\small{Am Campus 1, 3400 Klosterneuburg, Austria}}

\date{}

\maketitle

\begin{abstract}
We consider Ising models in two and three dimensions, with short range ferromagnetic and 
long range, power-law decaying, antiferromagnetic interactions. 
We let $J$ be the ratio between the strength of the ferromagnetic to antiferromagnetic interactions.
The competition between these two kinds of interactions induces the system to form 
domains of minus spins in a background of plus spins, or vice versa. 
If the decay exponent $p$ of the long range interaction is larger than $d+1$,
with $d$ the space dimension, 
this happens for all values of $J$ smaller than a critical value $J_c(p)$, beyond which the ground state is homogeneous. In this paper, we give a characterization of the infinite 
volume ground states of the system, for $p>2d$ and $J$ in 
a left neighborhood of $J_c(p)$. In particular, we prove that the quasi-one-dimensional states 
consisting of infinite stripes ($d=2$) or slabs ($d=3$), all of the same optimal width and orientation, and alternating magnetization, are infinite volume ground states. 
Our proof is based on localization bounds combined with reflection positivity.
\end{abstract}

\section{Introduction and main results}\label{sec1}

The problem of proving the emergence of periodic patterns in systems 
with competing interactions is ubiquitous in several areas of physics, biology and material science \cite{Ba}, 
ranging from superconductor physics \cite{EKT99}, micro-magne\-tism \cite{DMW00}, polymer suspensions \cite{HAC00}, martensitic phase transitions \cite{KM}, quantum Hall systems
\cite{FK99}, to metal-oxide-semiconductor field-effect transistors \cite{SK04}, nuclear matter
\cite{OMYT13}, and many others \cite{CDSN12,CN11}. In all these systems, experiments or simulations show evidence for the formation of 
remarkable patterns in suitable regions of the phase diagram. Examples are stripes, bubbles, zig-zag 
patterns, and columnar phases. The fundamental understanding of these phenomena is still in a primitive stage, mostly 
based on variational computations and special assumptions on the structure of the 
low-energy states. There are just a few special cases where the periodicity of the ground state can be proved from first principles \cite{FT15, HR80,KL86,S05,T06}.

A particularly interesting and poorly understood phenomenon is that of periodic stripe formation
\cite{MWRD95,MSN15,SW92,SS00}. In a series of papers, this phenomenon was studied in Ising and related models with short range attractive and long range repulsive interactions.
The method of block reflection positivity led   to 
rigorous proof the existence of periodic striped states both in one dimension \cite{GLL06,GLL09a,GLL09b} and in certain 
two-dimensional models, including a toy model model for martensitic phase  transitions \cite{GMu12}
and a model of in-plane spins with discrete orientations and dipolar interactions \cite{GLL07}. 
However, the physically 
interesting case of out-of-plane spins with dipolar, or dipolar-like, interactions, which is of great importance for the physics of thin magnetic films, eluded any rigorous treatment so far. 

In a recent work \cite{GLS}, we succeeded in computing the specific ground state energy of such a system, with 
power-law  decaying repulsive interactions and decay exponent $p>2d$ in $d=2,3$ space 
dimensions,
asymptotically as the ferromagnetic transition line is approached. Our estimates allowed us to 
prove emergence of periodic stripe order in a suitable asymptotic sense, but they were
not strong enough to fully control the ground state structure, or to prove breaking of rotational symmetry 
in the ground state. In this paper we extended the ideas of \cite{GLL11,GLS} and prove that periodic striped
states of optimal width are exact infinite volume ground states. Moreover we give a characterization of 
infinite volume ground states that are invariant under translations by one (for $d=2$) or two (for $d=3$) independent fixed lattice vectors.

The setting is the following: consider Ising models defined by the formal Hamiltonian
\be H(\ss)=-J\sum_{\media{\xx,\yy}}(\s_\xx\s_\yy-1)+\sum_{\{\xx,\yy\}}\frac{(\s_\xx\s_\yy-1)}{|\xx-\yy|^p}
\label{1.1}
\ee
where $\ss\in\{\pm1\}^{\mathbb Z^d}$, $d \geq 2$.  The first sum in \eqref{1.1} ranges over nearest neighbor pairs in $\mathbb Z^d$, while the second over pairs of distinct sites in $\mathbb Z^d$.
{For different values of the exponent $p$, this model is used
to describe the effects of frustration induced in magnetic films 
by the presence of
dipolar interactions ($p=3$) or in charged systems by the presence of an
unscreened Coulomb interaction ($p=1$), as well as many other frustrated systems
\cite{AWMD96,BCK07,CMST06,CDSN12,CN11,CEKNT96,CPPV09, CV89,EJ10,GTV00,
LEFK94,MWRD95,NBH08,OTC09,PC07,PGSBPV10,RRT06,SS00, VSPPP08}.}

{In this paper, we choose the exponent $p$ to satisfy the constraint $p>2d$. As discussed in a previous work} \cite{GLS},
if $J> J_c$, with 
\be J_c:=\sum_{y_1>0,\ \yy^\perp\in\mathbb
Z^{d-1}}\frac{y_1}{(y_1^2+|\yy^\perp|^2)^{p/2}}\;,\ee
then there are exactly two ground states, $\s_\xx\equiv +1$ $\forall
\xx\in\mathbb Z^d$, and $\s_\xx\equiv -1$
$\forall \xx\in\mathbb Z^d$. For $J<J_c$, the ground state is {\it not} uniform, and for $J$ close to $J_c$ it was 
conjectured to be a periodic striped configuration, i.e., a 
quasi-one-dimensional periodic configuration of the form $(\ss^{(h)})_{\xx}={\rm sign}(\sin(\p(x_1+1/2)/h))$,
or translations or rotations thereof, for a suitable stripe width $h\in \NNN$. In this paper, we prove this conjecture, 
and characterize the 
set of infinite volume ground states invariant under translations generated by $d-1$ lattice vectors.

For simplicity, we restrict the discussion to $d=2$ and $p>4$ from now on. Similar considerations are valid in 
$d=3$ and $p>6$ (or, in fact, for any $d\geq 2$ with $p>2d$). In Appendix \ref{appa.1} we explain 
how to adapt the proof to dimension three and higher. 
Let $e_{\rm s}(h)$ be the energy per site of $\ss^{(h)}$ computed via 
\eqref{1.1}. We let
$h^*={\rm argmin}_{h\in\mathbb N} e_{\rm s}(h)$, which is uniquely defined for almost all\footnote{There are exceptional values of $J$ for which
$e_{\rm s}(h)$ has two minimizers, $h^*$ and $h^*+1$.}
choices of $J$. We denote by $\underline\s^*=\ss^{(h^*)}\in\{\pm 1\}^{\mathbb Z^2}$, and we call it an
{\it optimal periodic striped configuration}. Other $4h^*-1$ optimal periodic striped configurations are obtained from $\ss^*$ via translations and rotations. 

In order to state our main result, we also need to introduce the following notions: the configuration $\underline s\in\{\pm 1\}^{\mathbb Z^2}$ is called an 
{\it infinite volume ground state} if it is energetically stable against compactly supported perturbations, 
that is, for any {\it finite} $X\subset \mathbb Z^2$, 
\be H_X(\ss_X|\underline s)-H_X(\underline s_X|\underline s)\ge 0,\quad \forall \ss_X\in\{\pm1\}^X\ee
where $\underline s_X$ is the restriction of $\underline s$ to $X$, and
\bea \nonumber H_X(\ss_X|\underline s)&=&-J\sum_{\substack{\media{\xx,\yy}:\\  \xx,\yy\in X}}( \s_\xx \s_\yy-1)+
\sum_{\substack{\{\xx,\yy\}:\\\xx,\yy\in X}}
\frac{(\s_\xx \s_\yy-1)}{|\xx-\yy|^p}\\
&&-J\sum_{\substack{\xx\in X,\ \yy\in X^c:\\ |\xx-\yy|=1}}(\s_\xx s_\yy-1)+
\sum_{{\xx\in X,\ \yy\in X^c}}
\frac{(\s_\xx s_\yy-1)}{|\xx-\yy|^p}.\eea
We shall say that two infinite volume ground states are {\it equivalent}, if they only differ on a finite set. 
The equivalence class of a given infinite volume ground state is called a {\it sector}. A sector is 
{\it trivial}, if it contains only one element. In terms of these notions, our main result can be summarized as follows.  

\begin{theorem}\label{thm1}
There exists $\e>0$ such that, if $J_c-\e<J<J_c$, then the optimal periodic striped configurations are 
infinite volume ground states, and their sectors are trivial. 
\end{theorem}

This result is a corollary of a quantitative lower bound on the energy of spin configurations, which will be 
formulated in Theorem \ref{thm3} below, after having introduced a few more definitions. Our quantitative bounds also 
allow us to characterize the infinite  volume ground states that are invariant under translation 
by a vector ${\bf n}\in\mathbb Z^2$. 

\begin{theorem}\label{thm2}
Under the same conditions as Theorem \ref{thm1}, any infinite volume ground state $\underline s$ that is invariant 
under translation by a vector ${\bf n}=(n_1,n_2)\in\mathbb Z^2$ is characterized by the following property: 
there exists an  ``interface'' of finite width, of the form $\mathcal I_{k_1,k_2}({\bf n})=\{\xx\in\mathbb Z^2: \ \xx\cdot {\bf n}_\perp\in[k_1,k_2]\}$, where ${\bf n}_\perp=(-n_2,n_1)$ and $k_1<k_2$ are two integers, such that 
$\underline s$ coincides with two of the optimal striped configurations on the two infinite components of $\big(\mathcal I_{k_1,k_2}({\bf n})\big)^c$.
\end{theorem}
Let us now introduce a few more 
definitions, which are required for the formulation of our quantitative lower bound on the energy 
of a generic spin configuration. 

\subsection{On good and bad}

\subsubsection{Contours and corners}\label{sec1.1.1}
\begin{figure}[h]
\centering
\includegraphics[width=.9\textwidth]{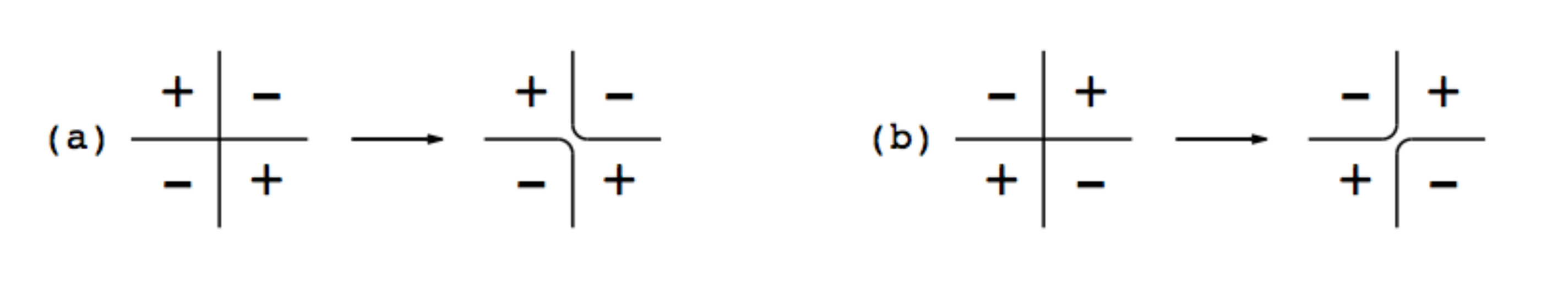}
\caption{In the case that 4 sides of the closed polygon $\G(\D)$ meet at a vertex $v$, we slightly deform 
$\G(\D)$ so that the two squares containing a $-$ spin become disconnected from the vertex itself. Case (a)
represents the situation where the minus spins are located at NE and SW of $v$, before and after the ``chopping".
Case (b)
represents the situation where the minus spins are located at NW and SE of $v$, before and after the ``chopping".
}
\label{fig.corner}
\end{figure}

Given $\ss\in\{\pm1\}^{\mathbb Z^2}$,
we let $\D=\{\xx\in\mathbb Z^2: (\ss)_\xx=-1\}$, 
and $\G(\D)$ be its boundary, i.e., the union of 
bonds of the dual lattice $(\mathbb Z^2)^*$ separating a point $\xx\in\D$ from a point $\yy\in\D^c$. 
At every vertex of 
$\G(\D)\cap(\mathbb Z^2)^*$, there can be either 2 
or 4 sides meeting.
In the case of 4 sides, we deform the polygon  
slightly by ``chopping off'' the edge from the squares containing a $-$ spin; see Fig.~\ref{fig.corner}.

After the chopping, $\G(\D)$ splits into disconnected polygons $\G_1,\G_2,\ldots,$ which are called
{\it contours}. The points where two orthogonal portions of a contour meet are called {\it corners}.
The 
sites of the dual lattice where a non-trivial chopping operation took place correspond to two corners. 
We denote by $N_c(\G_i)$ the number of corners of $\G_i$, and $N_c(\D)=\sum_{i}N_c(\G_i)$. 
Note that, if $\ss$ is a compactly supported perturbation of 
$\ss^*$, that is, if the set 
$\{\xx\in\mathbb Z^2: \ (\ss)_\xx\neq(\ss^*)_\xx\}$ is finite, then 
$N_c(\D)<+\infty$.

\subsubsection{Tiles}\label{sec1.1.2}
Given an integer $\ell$, we pave $\mathbb Z^2$ with tiles of side $\ell$ and, given a tile $T$, we denote by $\G_T(\D)$ 
the union of the bonds in $\G(\D)$ that separate a point in $\D\cap T$ from a point in $\D^c$. 
Note that this convention assigns uniquely every bond in $\G(\D)$ to one of the tiles. 
The connected components of $\G_T(\D)$ are contained in the contours $\G_1,\ldots$, 
and are denoted by $\G_{T,i}$, $i=1,\ldots,r_T$. Given a maximal straight portion of $\G_{T,i}$, we 
assign to each of its two endpoints a ``number of corners'', which can be either $1/2$ or  $0$, 
depending on whether or not the given endpoint 
coincides with one of the corners in $\cup_{j\ge 1}\G_j$. This assignment induces a notion
of ``number of corners in the tile $T$'', to be denoted by $n_c(T)$, which is the sum 
of the number of corners of all the endpoints of the straight portions of $\G_{T,i}$, with $i=1,\ldots, r_T$.
Note that $n_c(T)$ can be either integer or half-integer, and $\sum_T n_c(T)=N_c(\D)$.

\subsubsection{Bad tiles and good regions}\label{sec1.1.3}

We now identify the tiles $T_i$ such that either $n_c(T_i)>0$, or they contain a square of side 
$\ell/5$ completely contained in $\D$ or in $\D^c$, to be called {\it hole}. 
We call these tiles {\it bad}\,\footnote{We shall choose $\ell$ large compared to the optimal stripe width $h^*$, which explains why we expect a hole to be energetically unfavorable, hence {\it bad}.}, and we let $\mathcal N_B$ be their number. For later convenience, we
also let $\mathcal N_B^{\rm hole}$ be the number of bad tiles containing a hole. 
The connected 
components of the complement of $\cup_{i=1}^{\mathcal N_B}T_i$, are denoted by $G_i$,
$i=1,\ldots, \mathcal N_G$, and are called the {\it good regions}.
By construction, any of these connected components contains portions of contours in $\G(\D)$ that are 
all straight with the same orientation, and have no corners. We denote by $\G_{G_i}$ the union
of contours in $\G(\D)$ contained in $G_i$. If the elements of 
$\G_{G_i}$ are all vertical, we will say that $G_i$ is {\it vertically striped}, and {\it horizontally striped} otherwise. 
Consider a good region $G_i$ that is vertically (resp. horizontally) striped. We say that $R$ is a 
``rectangular portion of stripe'' in $G_i$, if $R$ is a rectangle completely contained in $\D$ or in $\D^c$,
with its two vertical (resp. horizontal) boundaries both belonging to $\G_{G_i}$. We also define 
the distance between its two vertical (resp. horizontal) boundaries to be the width of $R$. 
Finally, we let
$A_h(G_i)$ denote the area of the union of all the rectangular portions of stripes 
of width $h$ in $G_i$. Note that, if $\ss$ is a compactly supported perturbation of $\ss^*$,
then  $\sum_{i=1}^{\mathcal N_G}A_h(G_i)$ is finite, $\forall h\neq h^*$.

We are now in the position of stating our quantitative lower bound on the energy 
$H_X(\underline s_X|\ss^*)$ of a spin configuration $\underline s$ with $\ss^*$ boundary conditions.

\begin{theorem}\label{thm3}
There exist positive constants $C_0$, $C_1$, $\e$ such that, if $J_c-\e<J<J_c$ and $C_0 h^*\le \ell\le (C_0(J_c-J))^{-1}$, 
then for every $\underline s\in\{\pm1\}^{\mathbb  Z^2}$ and every finite set $X\subset \mathbb Z^2$,
\bea H_X(\underline s_X|\underline \s^*)&\ge& H_X(\ss^*_X|\ss^*)
+C_1\Big(N_c+(J_c-J)^{\frac{p-2}{p-3}}\ell^2\mathcal N_B^{\rm hole}\Big)\nonumber\\
&+&\frac12\sum_{h\neq h^*}\sum_{i=1}^{\mathcal N_G}(e_{\rm s}(h)-e_{\rm s}(h^*))
A_h(G_i),\label{1.6}\eea
where $N_c$, $\mathcal N_B^{\rm hole}$, and $G_i$ are, respectively, the number of corners, the
number of bad tiles containing a hole, and the good regions, associated with the infinite spin configuration $\ss=(\underline s_X,\ss^*_{X^c})$ coinciding with $\underline s_X$ on 
$X$ and with $\ss^*$ on $X^c$, defined via tiling with squares of side length $\ell$ as described above.
\end{theorem}

{\bf Remark.} Since $h^* \sim (J_c-J)^{-1/(p-3)} \ll (J_c-J)^{-1}$ for small $J_c-J$ (compare with Eq.~\eqref{app0.4} below), the condition on $\ell$ can be fulfilled for $J$ close to $J_c$. Note also that $e_{\rm s}(h^*) \sim (J_c-J)^{(p-2)/(p-3)}$ for small $J_c-J$, which agrees with the factor multiplying $ \mathcal N_B^{\rm hole}$ in the second term on the right side of \eqref{1.6}. 
\medskip

{\bf Remark.} The prefactor $1/2$ in the second line of \eqref{1.6} can be replaced by any number 
less than 1, at the expense of modifying the constants $C_0$ and $\e$. 

\medskip

Theorem \ref{thm3} implies, in particular, that $\ss^*$ is an infinite volume ground state, and that 
every state $\ss$ that is a compactly supported perturbation of it, is {\it not} a ground state, simply 
because any such state necessarily has corners and, therefore, by \eqref{1.6},
it has strictly larger energy than $\ss^*$. This immediately implies Theorem \ref{thm1}. 

In order to see that also Theorem \ref{thm2} is a consequence of Theorem \ref{thm3}, consider an
infinite volume ground state $\underline s$ that is invariant under translations by an integer vector 
${\bf n}$.  Let $\L\subset \mathbb Z^2$ be a square box of side $L$, and note that 
the energy price for changing the boundary conditions from $\underline s$ to $\ss^*$ scales like the boundary, that is 
\be \big|H_\L(\ss_\L|\underline s)-H_\L(\ss_\L|\ss^*)\big|\le 2(J+J_c)|\partial \L|\ee
for any $\ss_\L$. Using this inequality and the very definition of infinite volume ground state, we have
\bea && H_\L(\ss_\L^*|\ss^*)+2(J+J_c)|\partial\L|\ge H_\L(\ss_\L^*|\underline s)
\ge\\
&&\qquad \qquad\ge H_\L(\underline s_\L|\underline s)\ge H_\L(\underline s_\L|\ss^*)-2(J+J_c)|\partial\L|.\nonumber\eea
Now we apply Theorem \ref{thm3}, thus obtaining
\be C_1\Big(N_c+(J_c-J)^{\frac{p-2}{p-3}}\ell^2\mathcal N_B^{\rm hole}\Big)+\frac12\sum_{h\neq h^*}\sum_{i=1}^{\mathcal N_G}(e_{\rm s}(h)-e_{\rm s}(h^*))
A_h(G_i)\le 4(J+J_c)|\partial\L|,\ee
where $N_c, \mathcal N_B^{\rm hole}$ and $G_i$ refer to the configuration $(\underline s_\L,\ss^*_{\L^c})$. In particular, $N_c$ is bounded by (const.)$L$. Since every corner not at the boundary of $\L$
is repeated with period ${\bf n}$, there can be at most a finite number of them (modulo translations by ${\bf n}$) independently of $L$. 
This means that these corners are all contained in a finite strip $\mathcal I_{k_1,k_2}({\bf n})$,
as claimed in Theorem \ref{thm2}, with $k_1,k_2$ independent of $L$. 
Similarly, we can argue that the holes and the stripes of width different from $h^*$ are all contained 
in a finite strip $\mathcal I_{k_1,k_2}({\bf n})$. This concludes the proof of Theorem \ref{thm2}, in the 
case that $h^*$ is unique. As observed above, there are exceptional values of $J$ for which 
$e_{\rm s}(h)$ has two minimizers, $h^*$ and $h^*+1$. In these cases, the discussion above leaves 
open the possibility that on one of the connected components of $\big(\mathcal I_{k_1,k_2}({\bf n})\big)^c$ the 
stripes are not all of the same width. However, this cannot be the case: by proceeding as in 
\cite[Section III.D]{GLL06}, one can prove that each pair of neighboring stripes of widths $h^*, h^*+1$
gives an extra positive contribution to the energy per unit stripe length. Therefore, pairs 
of stripes of different widths are all contained in a finite strip $\mathcal I_{k_1,k_2}({\bf n})$, 
and Theorem \ref{thm2} follows. 

\bigskip

The rest of the paper is devoted to the proof of Theorem \ref{thm3}.

\section{Proof of Theorem \ref{thm3}}

The proof of Theorem \ref{thm3} is divided into several steps, and uses many notations and ideas introduced in \cite{GLS}, which will be recalled here. As a preliminary step, we reduce to plus boundary conditions, which allows us to use the droplet formulation for the energy as in \cite{GLS}, see \eqref{2.15} below. Once the energy is expressed in terms of droplets, we can localize the energy 
in the bad tiles and good regions, by proceeding in a way analogous to \cite{GLS}. 
The key technical novelty of this paper, as compared to \cite{GLS}, is an efficient way of estimating the 
energy in the good regions, which may have a complicated geometrical shape. The 
crucial estimate is summarized in Lemma \ref{lm2.2} below, whose proof is given in Section \ref{sec2.3}.

Before entering the proof, we recall some notation: we let $\t=2(J-J_c)$, which is assumed to be 
negative and small, in absolute value. We recall that $h^*$ is of the order $|\t|^{-1/(p-3)}$ and the 
specific energy of an optimal striped configuration, $e_{\rm s}(h^*)$, is negative and of the order 
$|\t|^{(p-2)/(p-3)}$ (compare with (\ref{app0.4}) below).

\subsection{Reduction to plus boundary conditions}

The $\ss^*$ boundary conditions, while natural in the perspective of proving uniqueness of the 
ground state, are not particularly convenient for using the droplet representation of \cite{GLS}. However, 
a few simple algebraic manipulations allow us to reduce to the same boundary conditions of \cite{GLS} 
(that is, plus boundary conditions) in a suitable enlarged box. To see this, rewrite $H_X(\ss_X|\ss^*)$ 
as 
\be H_X(\ss_X|\ss^*)=H_X(\ss^*_X|\ss^*)+\lim_{\L\nearrow\mathbb Z^2}
\big[H^{\rm per}_{\L}(\ss_X,\ss^*_{\L\setminus X})-H^{\rm per}_{\L}(\ss^*_{\L})\big]\label{2.1}
\ee
where $\L$ is a square box of side $L$, which we choose to be divisible by $2h^*$, 
$(\ss_X,\ss^*_{\L\setminus X})$ is the configuration on $\L$ whose restriction to $X\subset \L$ 
(resp. $\L\setminus X$) coincides with $\ss_X$ (resp. $\ss^*_{\L\setminus X}$), and 
$H^{\rm per}_\L(\ss_\L)=H_\L(\ss_\L|\ss^{\rm per}_\L)$
is the Hamiltonian with periodic, rather than $\ss^*$, boundary conditions (here $\ss_\L^{\rm per}$
is the periodic extension of $\ss_\L$ over $\mathbb  Z^2$).

Now, $H^{\rm per} _{\L}(\ss^*_{\L})=e_{\rm s}(h^*)|\L|$ and 
$H^{\rm per} _{\L}(\ss_X,\ss^*_{\L\setminus X})$ can be further rewritten in terms of a Hamiltonian with plus 
boundary conditions:
\be H^{\rm per} _{\L}(\ss_X,\ss^*_{\L\setminus X})=\lim_{M\to\infty}\frac1{M^2}H^+_{\L_M}
\big((\ss_X,\ss^*_{\L\setminus X})^{M^2}\big)\label{2.2}\ee
where $\L_M$ is a square box of side $LM$, to be thought of as the union of $M^2$ copies of $\L$, 
and $(\ss_X,\ss^*_{\L\setminus X})^{M^2}$ is a symbol for the configuration on $\L_M$ obtained by 
juxtaposing $M^2$ copies of 
$(\ss_X,\ss^*_{\L\setminus X})$, one in each of the copies of $\L$. Moreover, $H^+_X(\ss_X)=
H_X(\ss_X|\ss^+)$ indicates the 
Hamiltonian with plus boundary conditions (here $\ss^+$ is the uniform infinite spin configuration 
consisting of plus spins everywhere). Finally, for later reference, we introduce the shorthand 
${\underline u}_{\L_M}$ for  the spin configuration $(\ss_X,\ss^*_{\L\setminus X})^{M^2}$ on $\L_M$,
and $\underline u$ for the infinite one coinciding with ${\underline u}_{\L_M}$ on $\L_M$ and with $\ss^+$ on the complement. From now on we shall consider 
${\underline u}_{\L_M}$ and ${\underline u}$ fixed once and for all. 

\subsection{Localization}

\begin{figure}
\centering
\begin{tikzpicture}
	\begin{scope}[shift={(0,0)}]	
	
	\fill[gray!10] (0,0) -- (5,0) -- (5,5) -- (0,5) -- (0,0) -- cycle;
	
	\fill[gray!30]  (0,0)--(0.7,0)--(0.7,-1.1)--(2.6,-1.1)--(2.6,5.3)--(2.2,5.3)--(2.2,-0.6)--(1.5,-0.6)--(1.5,1)--(0,1)--(0,0)--cycle;
	\draw[very thick] (0,0)--(0.7,0)--(0.7,-1.1)--(2.6,-1.1)--(2.6,5.3)--(2.2,5.3)--(2.2,-0.6)--(1.5,-0.6)--(1.5,1)--(0,1)--(0,0)--cycle;

	\fill[gray!30] (3,1.2)--(4.2,1.2)--(4.2,5.4)--(4.8,5.4)--(4.8,5.8)--(3.2,5.8)--(3.2,3.8)--(3.6,3.8)--(3.6,5.4)--(3.9,5.4)--(3.9,2)--(3,2)--(3,1.2)--cycle;
	\draw[very thick] (3,1.2)--(4.2,1.2)--(4.2,5.4)--(4.8,5.4)--(4.8,5.8)--(3.2,5.8)--(3.2,3.8)--(3.6,3.8)--(3.6,5.4)--(3.9,5.4)--(3.9,2)--(3,2)--(3,1.2)--cycle;

	\fill[gray!30] (3.5,0.6)--(6,0.6)--(6,-1.1)--(3.5,-1.1)--(3.5,0.6)--cycle;
	\draw[very thick] (3.5,0.6)--(6,0.6)--(6,-1.1)--(3.5,-1.1)--(3.5,0.6)--cycle;

	\fill[gray!30] (1.8,2)--(1.8,3.2)--(1.3,3.2)--(1.3,2.7)--(0.8,2.7)--(0.8,2)--(1.8,2)--cycle;
	\draw[very thick] (1.8,2)--(1.8,3.2)--(1.3,3.2)--(1.3,2.7)--(0.8,2.7)--(0.8,2)--(1.8,2)--cycle;
	
	\fill[gray!30]  (-1.3,2.7)--(0.5,2.7)--(0.5,4.2)--(0,4.2)--(0,3.65)--(-1.3,3.65)--(-1.3,3.3)--(0,3.3)--(0,3)--(-1.3,3)--(-1.3,2.7)--cycle;
	\draw[very thick] (-1.3,2.7)--(0.5,2.7)--(0.5,4.2)--(0,4.2)--(0,3.65)--(-1.3,3.65)--(-1.3,3.3)--(0,3.3)--(0,3)--(-1.3,3)--(-1.3,2.7)--cycle;
	
	\fill[gray!30]  (0,5)--(0,6)--(1.8,6)--(1.8,4.5)--(1.,4.5)--(1.,5)--(0,5)--cycle;
	\draw[very thick] (0,5)--(0,6)--(1.8,6)--(1.8,4.5)--(1.,4.5)--(1.,5)--(0,5)--cycle;
	\draw[thick, dotted] (0,0) -- (5,0) -- (5,5) -- (0,5) -- (0,0) -- cycle;

	\end{scope}

\begin{scope}[shift={(8,0)}]	
	\fill[gray!10] (0,0) -- (5,0) -- (5,5) -- (0,5) -- (0,0) -- cycle;
	
	\fill[gray!30]  (1.5,0)--(0.,0)--(0,1)--(1.5,1)--(1.5,0)--cycle;
	\draw[very thick] (0.7,0)--(0.,0)--(0,1)--(1.5,1)--(1.5,0);
		\draw (0.7,.5) node {$\beta_7$};

	\fill[gray!30]  (2.2,0)--(2.6,0)--(2.6,5)--(2.2,5)--(2.2,0)--cycle;
		\draw[very thick] (2.6,0)--(2.6,5);
		\draw[very thick] (2.2,0)--(2.2,5);
		\draw (2.4,2.5) node {$\beta_2$};
	
	\fill[gray!30] (3,1.2)--(4.2,1.2)--(4.2,5)--(3.9,5)--(3.9,2)--(3,2)--(3,1.2)--cycle;
	\fill[gray!30] (3.2,5)--(3.2,3.8)--(3.6,3.8)--(3.6,5)--(3.2,5)--cycle;
	\draw[very thick] (3.9,5)--(3.9,2)--(3,2)--(3,1.2)--(4.2,1.2)--(4.2,5);
	\draw[very thick] (3.2,5)--(3.2,3.8)--(3.6,3.8)--(3.6,5);
	\draw (3.6,1.6) node {$\beta_4$};
	\draw (3.4,4.4) node {$\beta_3$};

	\fill[gray!30] (3.5,0.6)--(5,0.6)--(5,0)--(3.5,0)--(3.5,0.6)--cycle;
	\draw[very thick] (3.5,0)--(3.5,0.6)--(5,0.6);
	\draw (4.25,.3) node {$\beta_8$};
	
	\fill[gray!30] (1.8,2)--(1.8,3.2)--(1.3,3.2)--(1.3,2.7)--(0.8,2.7)--(0.8,2)--(1.8,2)--cycle;
	\draw[very thick] (1.8,2)--(1.8,3.2)--(1.3,3.2)--(1.3,2.7)--(0.8,2.7)--(0.8,2)--(1.8,2)--cycle;
	\draw (1.3,2.35) node {$\beta_6$};
	
	\fill[gray!30]  (0,2.7)--(0.5,2.7)--(0.5,4.2)--(0,4.2)--(0,2.7)--cycle;
	\draw[very thick] (0,2.7)--(0.5,2.7)--(0.5,4.2)--(0,4.2)--(0,3.65);
	\draw[very thick] (0,3)--(0,3.3);
	\draw (0.25,3.45) node {$\beta_5$};
	
	\fill[gray!30]  (1.8,5)--(1.8,4.5)--(1.,4.5)--(1.,5)--(1.8,5)--cycle;
	\draw[very thick] (1.8,5)--(1.8,4.5)--(1.,4.5)--(1.,5);
	\draw (1.4,4.75) node {$\beta_1$};
	
	\draw[thick, dotted] (0,0) -- (5,0) -- (5,5) -- (0,5) -- (0,0) -- cycle;

		\end{scope}
	
			\end{tikzpicture}
			\caption{On the left: a square region $Q$ (in light grey, with dotted boundary) and the 
			droplets having non-zero intersection with it. On the right: the corresponding bubble configuration $\mathcal B_Q$ after localization in $Q$. Note that a single droplet can 
			give rise after localization to more than one bubble: e.g., $\b_2$ and $\b_7$ originate from the same droplet, and similarly for $\b_3$ and $\b_4$. Note also that the contour of a bubble does not necessarily coincide with the boundary of its droplet: in general, it is contained in it, and may even be disconnected (as in the case of $\b_5$, whose contour consists of two disconnected portions).}
\label{fig2bistris}
\end{figure}
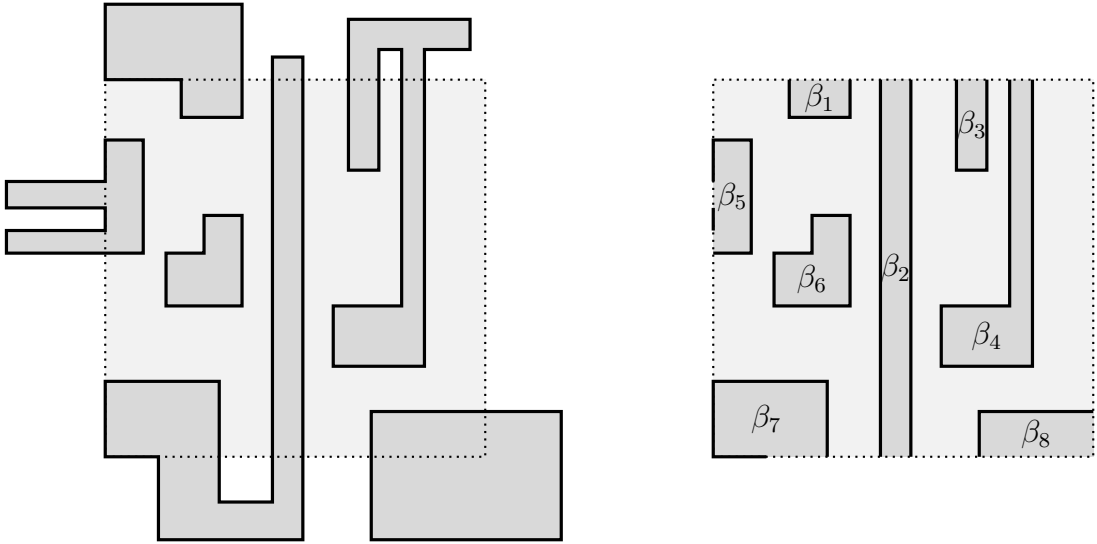

We now re-express $H^+_{\L_M}(\underline u_{\L_M})$ in terms of the droplets
representation introduced in \cite{GLS}. Using the notation introduced in Section \ref{sec1.1.1},
we let $\D$ be the region of minus spins associated with ${\underline u}$, $\G(\D)$ its boundary, and $\G_1,\ldots,\G_r$ the corresponding contours. We also denote by $\mathfrak G(\D)$ the collection 
of contours, $\mathfrak G(\D)=\{\G_1,\ldots,\G_r\}$.
Note that $\D$, $\G(\D)$, $\mathfrak G(\D)$ and $r$ are finite, because of the plus boundary conditions.  
As in \cite{GLS}, we denote by $\d_i$ the maximal connected components of $\D$, and by $\mathcal D(\D)$ their collection. Given $\d\in\mathcal D(\D)$, we also let $\G(\d)$ be the boundary of $\d$,
and $N_c(\d)$ its number of corners. 
In terms of these notations, we can re-express the energy 
of $\underline u_{\L_M}$ as
\begin{equation} H^+_{\L_M}(\underline u_{\L_M})=
2J\sum_{\G\in\mathfrak G(\D)}|\G|+
\sum_{\d\in\DD(\D)}
U(\d)+\frac12\sum_{\substack{\d,\d'\in\mathcal D(\D)\\ \d\neq\d'}}W(\d,\d')\;,\label{2.15}\end{equation}
where
\be U(\d):= -2\sum_{\xx\in\d}\sum_{\yy\in\mathbb Z^2\setminus \d}\frac1{|\xx-\yy|^p},\qquad 
W(\d,\d'):= 4\sum_{\xx\in\d}\sum_{\yy\in\d'} \frac1{|\xx-\yy|^p}\,.\ee
Let us consider the partition $\PP$ of $\L_M$ defined by the bad tiles and good 
regions of $\L_M$, in the sense of Section \ref{sec1.1.3}: $\mathcal P=\{T_i\}_{i=1}^{\mathcal N_B}\cup
\{G_i\}_{i=1}^{\mathcal N_G}$. We now localize the energy in the elements of $\mathcal P$,
by proceeding as in \cite[Section 3]{GLS}. More precisely, 
we derive a lower bound on the energy $H^+_{\L_M}(\underline u_{\L_M})$ in the form of a sum of local 
energies $E_Q(\mathcal B_Q)$, each depending only on the ``bubble configuration'' $\mathcal B_Q$ 
within the region $Q\in\mathcal P$. The notion of bubble configuration was introduced in 
\cite[Section 3]{GLS} and is recalled here: given $Q\in\mathcal P$ and $\d\in\mathcal D$, we denote by $\G_Q(\d)$ the portion of $\G(\d)$ belonging to $Q$. Moreover, if $\d_Q=\d\cap Q$, we define $\bar \d_Q^{(1)},\ldots, \bar \d_Q^{(m_Q(\d))}$ to be the maximal connected 
components of $\d_Q$, and $\bar\G_Q^{(1)},\ldots, \bar\G_Q^{(m_Q(\d))}$ to be the portions of $\G_Q(\d)$ belonging to 
the boundary of $\bar\d_Q^{(1)},\ldots, \bar\d_Q^{(m_Q(\d))}$, respectively. 
We shall refer to the pair $(\bar\d_Q^{(i)},
\bar\G_Q^{(i)})$ as to a {\it bubble}
in $Q$ originating from $\d$. We shall indicate by ${\mathcal B}_Q(\d)$ the set of bubbles in $Q$ originating from 
$\d$, and by ${\mathcal B}_Q=\cup_{\d\in\mathcal D}{\mathcal B}_Q(\d)$ the total set of bubbles in $Q$. (See Fig.~\ref{fig2bistris}.)

Given a bubble $\b=(\d_\b,\G_\b)\in\mathcal B_Q$, we also define its localized self-energy as
\be u_Q(\b)=-\sum_{b\in \G_\b}
\ \sum_{{\Vn\neq\V0}}\frac{\min\{|n_1|,d^{Q}_b(\d_\b)\}}{|\Vn|^p},\ee
where $d_b^Q(\d_\b)$ is the distance between $b$ and the bond $b'\in\G_\b$ facing it\footnote{
The notion of ``bond facing $b$ in $\b$'' is defined as follows. Let us suppose for definiteness that $b\in\G_\b$ is vertical and that it separates a point $\xx_b\in\d_\b$ 
on its immediate right from a point $\yy_b=\xx_b-(1,0)\not\in\d_\b$ on its immediate left. 
We say that $b$ faces $b'\in\G_\b$, and vice versa, if: (i) $b'$ is vertical; (ii) $b'$ separates a point $\xx_{b'}\in\d_\b$ 
on its immediate left from a point $\yy_{b'}=\xx_{b'}+(1,0)\not\in\d_\b$ on its immediate right; (iii) the points $\xx_b$ and 
$\xx_{b'}$ are at the same height, i.e., $[\xx_b]_2=[\xx_{b'}]_2$, and all the points on the same row between them 
belong to $\d_\b$: in other words, $\xx_b+(j,0)\in\d_\b$, for all $j=0,\ldots,[\xx_{b'}]_1-[\xx_{b}]_1$. An analogous definition is valid for horizontal bonds. If $b$ does not face any $b'\in \G_\b$, we say that 
$b$ faces the boundary of $Q$.} in $\b$, 
if there is such a bond, and is infinite, if $b$ faces the boundary of $Q$.
In terms of these notions, our localization bound takes the following form: 
\be H^+_{\L_M}(\underline u_{\L_M})\ge \sum_{i=1}^{\mathcal N_B}E_{T_i}(\mathcal B_{T_i})+
\sum_{i=1}^{\mathcal N_G}E_{G_i}(\mathcal B_{G_i}),\label{2.7}\ee
where
\bea && \hskip-1.1truecm E_{T_i}(\mathcal B_{T_i})=\sum_{\b\in\mathcal B_{T_i}}
[2J|\G_\b|+u_{T_i}(\b)]+\frac12\sum_{\substack{\b,\b'\in\mathcal B_{T_i},\\ \b\neq\b'}}W(\d_\b,\d_{\b'})+2^{1-p/2}n_c(T_i),\quad \label{eq:2.12}\\
&&\hskip-1.1truecm E_{G_i}(\mathcal B_{G_i})=\sum_{\b\in\mathcal B_{G_i}}
[2J|\G_\b|+u_{G_i}(\b)]+\frac12\sum_{\substack{\b,\b'\in\mathcal B_{G_i},\\ \b\neq\b'}}^*W(\d_\b,\d_{\b'}),
\label{eq2.8}\eea
and the $*$ on the last sum indicates the following constraint: if the bubbles in $G_i$ are all vertical 
(resp. horizontal) we only sum over pairs of bubbles that do not 
overlap after arbitrary translations in the vertical (resp. horizontal) direction. 

\medskip

{\bf Proof of \eqref{2.7}.} We start from \eqref{2.15}. The goal is to bound it from below by 
a sum of terms, each of which is localized in an element $Q$ of $\mathcal P$. The first term on 
the right side of \eqref{2.15} is already local, i.e., it can be rewritten exactly as 
$2J\sum_{Q\in \mathcal P}\sum_{\b\in\mathcal B_Q}|\G_\b|$, which leads to the corresponding terms in 
\eqref{eq:2.12} and \eqref{eq2.8}. The interaction $W(\d,\d')$ on 
the right side of \eqref{2.15} can be rewritten as 
$$W(\d,\d')=\sum_{Q,Q'\in \mathcal P}
\sum_{\substack {\b\in\mathcal B_Q(\d)\\ 
\b'\in\mathcal B_{Q'}(\d')}}W(\d_\b,\d_{\b'}),
$$
which is bounded from below by dropping the terms with $Q\neq Q'$ (recall that the interaction is positive), so that 
\be\frac12\sum_{\substack{\d,\d'\in\mathcal D(\D)\\ \d\neq\d'}}W(\d,\d')\ge 
\frac12 \sum_{Q\in \mathcal P}
\sum_{\substack {\b,\b'\in\mathcal B_Q\\ 
\b\neq\b'}}^{**}W(\d_\b,\d_{\b'}),
\label{2.9-1}\ee
and the $**$ on the sum indicates the constraint that the two droplets $\d,\d'$ in $ \mathcal D(\D)$,
which $\d_\b,\d_{\b'}$ belong to, are different from each other, $\d\neq\d'$.

Regarding the second term on the right side of \eqref{2.15}, we bound it from below by using 
 \cite[Eq.(2.9)]{GLS}, that is 
\be U(\d)\ge -\sum_{b\in\G(\d)}
\ \sum_{{\Vn\neq\V0}}\frac{\min\{|n_1|,d_b(\d)\}}{|\Vn|^p}+2^{1-\frac{p}2}N_c(\d)+4\sum_{\{\xx,\yy\}\in\mathcal P(\d)}
\frac1{|\xx-\yy|^p}\;,\label{eq:2.9}
\ee
where $d_b(\d)$ is the distance between $b$ and the bond $b'$ facing it in $\d$, and $\mathcal P(\d)$ is the set of unordered pairs of distinct sites in $\d$ such that both 
$\mathcal C^{hv}_{\xx\to\yy}$ and $\mathcal C^{vh}_{\xx\to\yy}$ cross at least two bonds of $\G(\d)$. 
Here $\mathcal C^{hv}_{\xx\to\yy}$ is the path on the lattice that goes from $\xx$ to $\yy$ consisting of two segments,
the first horizontal and the second vertical. Similarly, $\mathcal C^{vh}_{\xx\to\yy}$ is the path on the lattice that goes from $\xx$ to 
$\yy$ consisting of two segments, the first vertical and the second horizontal (note that the two paths can coincide, in the case that 
$x_i=y_i$ for some $i\in\{1,2\}$). 

The first term on the right side of \eqref{eq:2.9} can be bounded from below as
\be -\sum_{b\in\G(\d)}
\ \sum_{{\Vn\neq\V0}}\frac{\min\{|n_1|,d_b(\d)\}}{|\Vn|^p}\ge \sum_{Q\in\mathcal D(\D)}
\sum_{\b\in\mathcal B_Q(\d)}u_Q(\b), \ee
which, after summation over $\d$, leads to the terms $\sum_{\b\in\mathcal B_Q}u_Q(\b)$ in 
\eqref{eq:2.12} and \eqref{eq2.8}. The second term on the right side of \eqref{eq:2.9} is local and, after summation over $\d$, can be rewritten as $2^{1-\frac{p}{2}}\sum_{i=1}^{\mathcal N_B}n_c(T_i)$ (recall that the good regions have no corners), which leads to the last term on the right side of \eqref{eq:2.12}.
Moreover, the sum over $\d$ of the last term on the right side of \eqref{eq:2.9} can be bounded from below by an expression similar to the right side of \eqref{2.9-1}, namely
\be \sum_{\d\in\mathcal D(\D)}4\sum_{\{\xx,\yy\}\in\mathcal P(\d)}\frac1{|\xx-\yy|^p}\ge
\frac12 \sum_{Q\in \mathcal P}
\sum_{\substack {\b,\b'\in\mathcal B_Q\\ 
\b\neq\b'}}^{\dagger}W(\d_\b,\d_{\b'}),\label{eq:2.12b}\ee
where the $\dagger$ on the sum indicates the constraint that $\d_\b,\d_{\b'}$ belong to the same droplet 
$\d\in\mathcal D(\D)$, and all the pairs of points $(\xx,\yy)$ in $\d_\b\times\d_{\b'}$ are such that 
$\{\xx,\yy\}\in\mathcal P(\d)$. Combining the right sides of \eqref{2.9-1} and \eqref{eq:2.12b} we 
obtain 
$$\frac12\sum_{Q\in \mathcal P}
\sum_{\substack {\b,\b'\in\mathcal B_Q\\ 
\b\neq\b'}}^{\dagger\dagger}W(\d_\b,\d_{\b'}),$$
where  the ${\dagger\dagger}$ on the sum indicates the constraint that: either $\d_\b,\d_{\b'}$ belong to different droplets in $\mathcal D(\D)$, or, if they belong to the same droplet in $\d$, they are such that all 
pairs of points $(\xx,\yy)$ in $\d_\b\times\d_{\b'}$ are in $\mathcal P(\d)$. Finally, note that: 
if $Q$ is a bad tile (which is a convex region), then the constraint ${\dagger\dagger}$ is automatically realized (i.e., it can be dropped),
which leads to the second term on the right side of \eqref{eq:2.12}; if $Q$ is a good region, then 
the constraint ${\dagger\dagger}$ is easily seen to be weaker than the one indicated by $*$ in \eqref{eq2.8}, which leads to the 
last term on the right side of \eqref{eq2.8}. \qed

\subsection{Lower bounds on the localized energies}

In this subsection we state the key lower bounds on the localized energies in the bad and good regions, 
and prove that they imply Theorem \ref{thm3}. Recall the definitions of $\mathcal N_B$, $\mathcal N_B^{\rm hole}$ and $A_h(G_i)$ given in Section \ref{sec1.1.3}. Recall also that $\ell$ is the side length of the tiles,
which enters the definition of the partition $\mathcal P$. 

\begin{lemma}\label{lm2.2}
There exist positive constants $c_0$, $c_1$ and $\e$ such that, if $-\e<\t<0$ and $\ell\ge c_0h^*$, then the energy $E_G$ of any good region $G\in\mathcal P$ satisfies
\be E_{G}({\mathcal B}_{G})\ge  e_{\rm s}(h^*)|G|-c_1|\t| |\partial G|+\frac12
\sum_{h\neq h^*} (e_{\rm s}(h)-e_{\rm s}(h^*))A_h(G).\label{eq:2.8}\ee
\end{lemma}

\begin{lemma}\label{lm1}
There exist  positive constants $c_0$, $c_2$ and $\e$ such that, if $-\e<\t<0$ and $c_0h^*\le \ell\le (c_0|\t|)^{-1}$, then the energy $E_T$ of any bad tile $T\in\mathcal P$ satisfies
\be E_{T}({\mathcal B}_{T})\ge \ell^2 e_{\rm s}(h^*)+c_2\big[n_c(T)+|\t|^{(p-2)/(p-3)}\ell^2\c_{\rm hole}(T)
\big],
\label{2.5bis}\ee
where $\c_{\rm hole}(T)$ is equal to 1 if $T$ contains a hole, and 0 otherwise.  
\end{lemma}

\bigskip

Lemma \ref{lm2.2} is  one of the main technical novelties of this 
paper, and its proof is described in detail in the next section. The proof of Lemma \ref{lm1} is simpler, 
it is an  extension of the bounds worked out in \cite{GLS}, and its proof 
is postponed to Section \ref{app10}. 

Combining the two lemmas, we can easily derive Theorem \ref{thm3}.
In fact, note that every portion of the boundary $\dpr G_i$ of a good region $G_i\in\mathcal P$ is 
adjacent to a bad tile, so that $\sum_{i=1}^{\mathcal N_G} |\partial G_i|\le 4\ell {\mathcal N}_B$.
Using this observation, together with the fact that every bad tile either has a positive number of corners (at least $1/2$) or a hole, and plugging \eqref{eq:2.8} and \eqref{2.5bis} into \eqref{2.7}, we obtain 
\bea  H^+_{\L_M}(\underline u_{\L_M}) &\ge &
e_{\rm s}(h^*)|\L_M|+\frac{c_2}2\Big(N_c(\D)+|\t|^{(p-2)/(p-3)}\ell^2\mathcal N_B^{\rm hole}\Big) \nonumber
\\ \label{2.12}
&&+\frac12
\sum_{h\neq h^*} \sum_{i=1}^{\mathcal N_G} (e_{\rm s}(h)-e_{\rm s}(h^*))A_h(G_i)\\ && +{\mathcal N}_B\Big(\frac{c_2}2\min\Big\{\frac12,|\t|^{\frac{p-2}{p-3}}\ell^2\Big\}-4c_1|\t|\ell\Big).
\nonumber\eea
Now, if 
$$\frac{8c_1}{c_2}|\t|^{-1/(p-3)}<\ell <\frac{c_2}{16 c_1}|\t|^{-1},$$ 
then the expression in parentheses that multiplies $\mathcal N_B$ is positive, and we can drop it for the purpose of a lower bound. Via the use of \eqref{2.1}--\eqref{2.2}, we thus arrive at Eq.~\eqref{1.6}.

\subsection{Proof of Lemma \ref{lm2.2}}\label{sec2.3}

\subsubsection{Deforming the good regions}

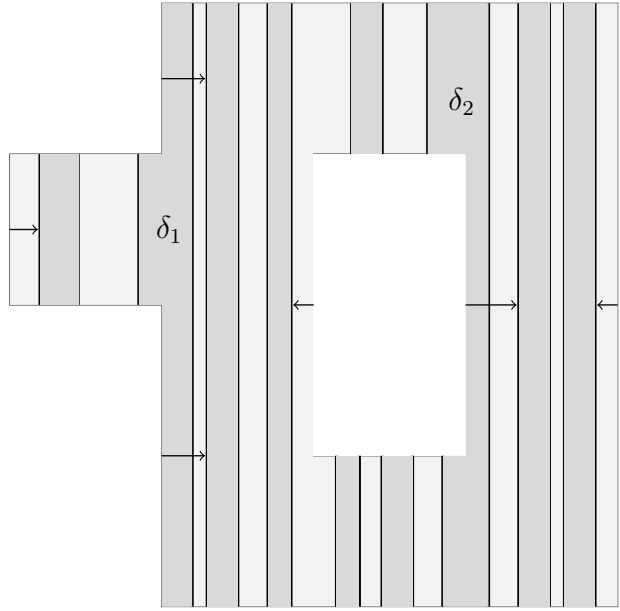
\begin{figure}
\centering
\begin{tikzpicture}
	\coordinate (P1) at (2,0); 
	\coordinate (P2) at (8,0); 
	\coordinate (P3) at (8,8); 
	\coordinate (P4) at (2,8); 
	\coordinate (P5) at (2,6);
	\coordinate (P6) at (0,6);  
	\coordinate (P7) at (0,4);
	\coordinate (P8) at (2,4);
	\coordinate (P9) at (4,2);
	\coordinate (P10) at (6,2);    
	\coordinate (P11) at (6,6);    
	\coordinate (P12) at (4,6);    
	
	\draw[very thin] (P1) -- (P2) -- (P3) -- (P4) -- (P5) -- (P6) -- (P7) -- (P8) -- (P1) -- cycle;

	\fill[gray!10] (P1) -- (P2) -- (P3) -- (P4) -- (P5) -- (P6) -- (P7) -- (P8) -- (P1) -- cycle;
	
	\draw[very thin] (P9) -- (P10) -- (P11) -- (P12) -- (P9)  -- cycle;

	\fill[white] (P9) -- (P10) -- (P11) -- (P12) -- (P9)  -- cycle;
	
	\coordinate (A1) at (0.4,4);
	\coordinate (A2) at (0.9,4);
	\coordinate (A3) at (1.7,4);
	\coordinate (A4) at (0.4,6);
	\coordinate (A5) at (0.9,6);
	\coordinate (A6) at (1.7,6);
         \coordinate (A7) at (2.4,0);
	\coordinate (A8) at (2.4,8);
  \coordinate (A9) at (2.6,0);
	\coordinate (A10) at (2.6,8);
 \coordinate (A11) at (3,0);
	\coordinate (A12) at (3,8);
 \coordinate (A13) at (3.4,0);
	\coordinate (A14) at (3.4,8);
 \coordinate (A15) at (3.7,0);
	\coordinate (A16) at (3.7,8);

	\draw[very thick] (A1)--(A4);
	\draw[very thick] (A2)--(A5);
	\draw[very thick] (A3)--(A6);

	\fill[gray!30] (A1) -- (A2) -- (A5) -- (A4) -- (A1)  -- cycle;
	
	\draw[very thick] (A7)--(A8);

	\fill[gray!30] (P1) -- (A7) -- (A8) -- (P4) -- (P5) -- (A6) -- (A3) -- (P8) -- (P1) -- cycle;
	
	\draw[very thick] (A9)--(A10);
	\draw[very thick] (A11)--(A12);
	\draw[very thick] (A13)--(A14);
\draw[very thick] (A15)--(A16);

	\fill[gray!30] (A9) -- (A11) -- (A12) -- (A10) -- (A9)  -- cycle;
	\fill[gray!30] (A13) -- (A15) -- (A16) -- (A14) -- (A13)  -- cycle;
	
	   \coordinate (B1) at (4.3,0);
	\coordinate (B2) at (4.3,2);
  \coordinate (B3) at (4.6,0);
	\coordinate (B4) at (4.6,2);
 \coordinate (B5) at (4.9,0);
	\coordinate (B6) at (4.9,2);
 \coordinate (B7) at (5.3,0);
	\coordinate (B8) at (5.3,2);
 \coordinate (B9) at (5.7,0);
	\coordinate (B10) at (5.7,2);
 \coordinate (B11) at (6.3,0);
	\coordinate (B12) at (6.3,8);

	\draw[very thick] (B1)--(B2);
	\draw[very thick] (B3)--(B4);
	\draw[very thick] (B5)--(B6);
\draw[very thick] (B7)--(B8);
\draw[very thick] (B9)--(B10);
\draw[very thick] (B11)--(B12);

	  \coordinate (C1) at (4.5,6);
	\coordinate (C2) at (4.5,8);
  \coordinate (C3) at (4.9,6);
	\coordinate (C4) at (4.9,8);
 \coordinate (C5) at (5.5,6);
	\coordinate (C6) at (5.5,8);

	\draw[very thick] (C1)--(C2);
	\draw[very thick] (C3)--(C4);
	\draw[very thick] (C5)--(C6);

	\fill[gray!30] (C1) -- (C3) -- (C4) -- (C2) -- (C1)  -- cycle;
	\fill[gray!30] (B1) -- (B3) -- (B4) -- (B2) -- (B1)  -- cycle;
\fill[gray!30] (B5) -- (B7) -- (B8) -- (B6) -- (B5)  -- cycle;
	\fill[gray!30] (B9) -- (B11) -- (B12) -- (C6) -- (C5) -- (P11) -- (P10) -- (B10) --(B9) -- cycle;

	  \coordinate (D1) at (6.7,0);
	\coordinate (D2) at (6.7,8);
  \coordinate (D3) at (7.1,0);
	\coordinate (D4) at (7.1,8);
 
 	  \coordinate (E1) at (7.3,0);
	\coordinate (E2) at (7.3,8);
  \coordinate (E3) at (7.7,0);
	\coordinate (E4) at (7.7,8);
 
	\draw[very thick] (D1)--(D2);
	\draw[very thick] (D3)--(D4);
	\draw[very thick] (E1)--(E2);
	\draw[very thick] (E3)--(E4);

	\fill[gray!30] (D1) -- (D3) -- (D4) -- (D2) -- (D1)  -- cycle;
	\fill[gray!30] (E1) -- (E3) -- (E4) -- (E2) -- (E1)  -- cycle;
\draw[arrows=->,line width=.5pt](0,5)--(0.37,5);
\draw[arrows=->,line width=.5pt](2,7)--(2.57,7);
\draw[arrows=->,line width=.5pt](2,2)--(2.57,2);
\draw[arrows=->,line width=.5pt](4,4)--(3.73,4);
\draw[arrows=->,line width=.5pt](6,4)--(6.67,4);
\draw[arrows=->,line width=.5pt](8,4)--(7.73,4);
\draw (2.1,5) node {$\delta_1$};
\draw (5.95,6.7) node {$\delta_2$};

		\end{tikzpicture}

\caption{A good region with its bubble configuration. The regions in dark grey are droplets, while those in light grey are their complement.
Note that not all of the droplets are rectangular, due to boundary effects. E.g., the droplets $\d_1$ and $\d_2$ are not rectangular. The arrows indicate the direction in which the vertical boundary segments move
under the deformation described in the text.}\label{fig2}
\end{figure}

We now discuss the proof of Lemma \ref{lm2.2}. Let us consider a good region $G\in\mathcal P$,
and let us assume 
without loss of generality that it only contains vertical stripes. We first want to slightly deform the domain 
$G$ and correspondingly change the bubble configuration within, in order to make 
all the bubbles rectangular: here we call  ``rectangular" a bubble $\b=(\d_\b,\G_\b)$ such that $\d_\b$ is a rectangle and $\G_\b$ is the 
union of its two vertical sides. 
Note that, in general, not all the bubbles in $G$ are rectangular, due to boundary effects (in fact, the boundary can 
partially ``cut" 
a portion of the rectangle, without disconnecting it, see e.g. the droplets $\d_1$ and $\d_2$ in Fig.~\ref{fig2}). 

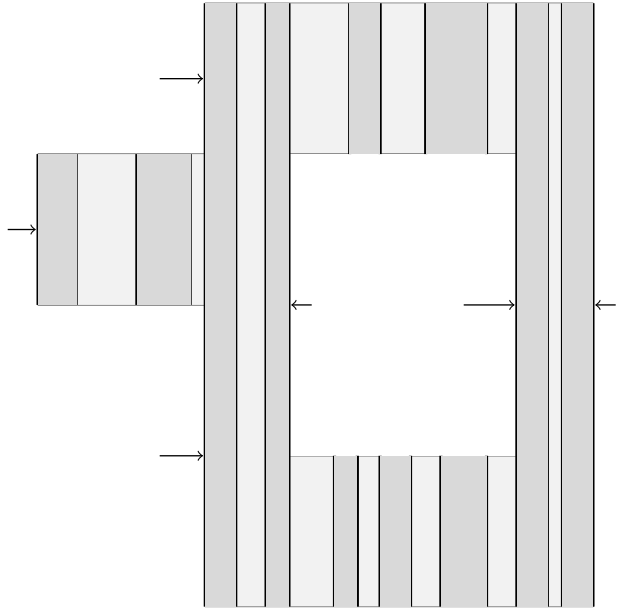
\begin{figure}
\centering
\begin{tikzpicture}
	\coordinate (P1) at (2,0); 
	\coordinate (P2) at (8,0); 
	\coordinate (P3) at (8,8); 
	\coordinate (P4) at (2,8); 
	\coordinate (P5) at (2,6);
	\coordinate (P6) at (0,6);  
	\coordinate (P7) at (0,4);
	\coordinate (P8) at (2,4);
	\coordinate (P9) at (4,2);
	\coordinate (P10) at (6,2);    
	\coordinate (P11) at (6,6);    
	\coordinate (P12) at (4,6);    
	
	\draw[very thin] (A9) -- (E3) -- (E4) -- (2.6,8) -- (2.6,6) -- (A4) -- (A1) -- (2.6,4) -- (A9) -- cycle;

	\fill[gray!10] (A9) -- (E3) -- (E4) -- (2.6,8) -- (2.6,6) -- (A4) -- (A1) -- (2.6,4) -- (A9) -- cycle;
	
	\draw[very thin] (3.7,2) -- (6.7,2) -- (6.7,6) -- (3.7,6) -- (3.7,2)  -- cycle;

	\fill[white] (3.7,2) -- (6.7,2) -- (6.7,6) -- (3.7,6) -- (3.7,2)  -- cycle;
	 
	\coordinate (A1) at (0.4,4);
	\coordinate (A2) at (0.9,4);
	\coordinate (A3) at (1.7,4);
	\coordinate (A4) at (0.4,6);
	\coordinate (A5) at (0.9,6);
	\coordinate (A6) at (1.7,6);
         \coordinate (A7) at (2.4,0);
	\coordinate (A8) at (2.4,8);
  \coordinate (A9) at (2.6,0);
	\coordinate (A10) at (2.6,8);
 \coordinate (A11) at (3,0);
	\coordinate (A12) at (3,8);
 \coordinate (A13) at (3.4,0);
	\coordinate (A14) at (3.4,8);
 \coordinate (A15) at (3.7,0);
	\coordinate (A16) at (3.7,8);

	\draw[very thick] (A1)--(A4);
	\draw[very thick] (A2)--(A5);
	\draw[very thick] (A3)--(A6);
	\draw[very thick] (2.4,4)--(2.4,6);

	\fill[gray!30] (A1) -- (A2) -- (A5) -- (A4) -- (A1)  -- cycle;
	
	\fill[gray!30] (A3) -- (2.4,4) -- (2.4,6) -- (A6) -- (A3)  -- cycle;
	
	\draw[very thick] (A9)--(A10);
	\draw[very thick] (A11)--(A12);
	\draw[very thick] (A13)--(A14);
\draw[very thick] (A15)--(A16);

	\fill[gray!30] (A9) -- (A11) -- (A12) -- (A10) -- (A9)  -- cycle;
	\fill[gray!30] (A13) -- (A15) -- (A16) -- (A14) -- (A13)  -- cycle;
	
	   \coordinate (B1) at (4.3,0);
	\coordinate (B2) at (4.3,2);
  \coordinate (B3) at (4.6,0);
	\coordinate (B4) at (4.6,2);
 \coordinate (B5) at (4.9,0);
	\coordinate (B6) at (4.9,2);
 \coordinate (B7) at (5.3,0);
	\coordinate (B8) at (5.3,2);
 \coordinate (B9) at (5.7,0);
	\coordinate (B10) at (5.7,2);
 \coordinate (B11) at (6.3,0);
	\coordinate (B12) at (6.3,8);

	\draw[very thick] (B1)--(B2);
	\draw[very thick] (B3)--(B4);
	\draw[very thick] (B5)--(B6);
\draw[very thick] (B7)--(B8);
\draw[very thick] (B9)--(B10);
\draw[very thick] (6.3,0)--(6.3,2);
\draw[very thick] (6.3,6)--(6.3,8);

	  \coordinate (C1) at (4.5,6);
	\coordinate (C2) at (4.5,8);
  \coordinate (C3) at (4.9,6);
	\coordinate (C4) at (4.9,8);
 \coordinate (C5) at (5.5,6);
	\coordinate (C6) at (5.5,8);

	\draw[very thick] (C1)--(C2);
	\draw[very thick] (C3)--(C4);
	\draw[very thick] (C5)--(C6);

	\fill[gray!30] (C1) -- (C3) -- (C4) -- (C2) -- (C1)  -- cycle;
	\fill[gray!30] (B1) -- (B3) -- (B4) -- (B2) -- (B1)  -- cycle;
         \fill[gray!30] (B5) -- (B7) -- (B8) -- (B6) -- (B5)  -- cycle;
         \fill[gray!30] (5.7,0) -- (6.3,0) -- (6.3,2) -- (5.7,2) -- (5.7,0)  -- cycle;
         \fill[gray!30] (5.5,6) -- (6.3,6) -- (6.3,8) -- (5.5,8) -- (5.5,6)  -- cycle;
         
	  \coordinate (D1) at (6.7,0);
	\coordinate (D2) at (6.7,8);
  \coordinate (D3) at (7.1,0);
	\coordinate (D4) at (7.1,8);
 
 	  \coordinate (E1) at (7.3,0);
	\coordinate (E2) at (7.3,8);
  \coordinate (E3) at (7.7,0);
	\coordinate (E4) at (7.7,8);
 
	\draw[very thick] (D1)--(D2);
	\draw[very thick] (D3)--(D4);
	\draw[very thick] (E1)--(E2);
	\draw[very thick] (E3)--(E4);

	\fill[gray!30] (D1) -- (D3) -- (D4) -- (D2) -- (D1)  -- cycle;
	\fill[gray!30] (E1) -- (E3) -- (E4) -- (E2) -- (E1)  -- cycle;
\draw[arrows=->,line width=.5pt](0,5)--(0.37,5);
\draw[arrows=->,line width=.5pt](2,7)--(2.57,7);
\draw[arrows=->,line width=.5pt](2,2)--(2.57,2);
\draw[arrows=->,line width=.5pt](4,4)--(3.73,4);
\draw[arrows=->,line width=.5pt](6,4)--(6.67,4);
\draw[arrows=->,line width=.5pt](8,4)--(7.73,4);

		\end{tikzpicture}
\caption{The deformed good region corresponding to the good region of Fig.~\ref{fig2}, after the moves
of the vertical boundary segments indicated by the arrows. Note that after the deformation all the droplets are rectangular, and all the connected components of the complement of the droplets (the 
connected light grey regions) are rectangular as well.}\label{fig3}
\end{figure}

In order to describe the deformation of $G$, think of the vertical boundary of $G$ as a 
union of segments $S_i$ of length $\ell$, induced by the tiling described in Sections \ref{sec1.1.2} and \ref{sec1.1.3}. 
By construction, every boundary segment $S_i$ faces a portion $\tilde S_i$ of length $\ell$ of the boundary of a {\em rectangular} bubble
in $\mathcal B_G$ that is closer than $2\ell/5$ to $S_i$ itself. We now deform the boundary $\partial G$ continuously, 
by moving the segments $S_i$ towards the 
interior of $G$, in such a way that they coincide with
$\tilde S_i$;  see Fig.~\ref{fig3}.

In this way we increase the boundary 
of $G$ by at most $|\partial G|$ itself, and we increase the energy by at most $|\t|\cdot|\partial G|$, which is 
acceptable for our purposes. We shall denote by $G'$ the new region obtained from $G$ by the 
deformations we just described, and the new bubble configuration $\mathcal B_{G'}$. We have
\be E_{G}({\mathcal B}_{G})\ge E_{G'}({\mathcal B}_{G'}) - |\tau|\cdot|\partial G|\,,
\ee
as just argued.

\subsubsection{Slicing and bounding the energy of the good region}\label{sec2.4.2}

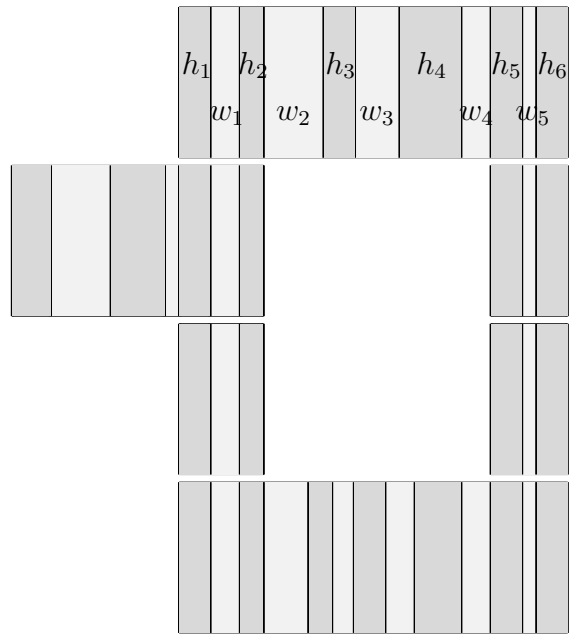
\begin{figure}
\centering
\begin{tikzpicture}
	\coordinate (P1) at (2,0); 
	\coordinate (P2) at (8,0); 
	\coordinate (P3) at (8,8); 
	\coordinate (P4) at (2,8); 
	\coordinate (P5) at (2,6);
	\coordinate (P6) at (0,6);  
	\coordinate (P7) at (0,4);
	\coordinate (P8) at (2,4);
	\coordinate (P9) at (4,2);
	\coordinate (P10) at (6,2);    
	\coordinate (P11) at (6,6);    
	\coordinate (P12) at (4,6);    
	
	\coordinate (A1) at (0.4,4);
	\coordinate (A2) at (0.9,4);
	\coordinate (A3) at (1.7,4);
	\coordinate (A4) at (0.4,6);
	\coordinate (A5) at (0.9,6);
	\coordinate (A6) at (1.7,6);
         \coordinate (A7) at (2.4,0);
	\coordinate (A8) at (2.4,8);
	\coordinate (A9) at (2.6,0);
	\coordinate (A10) at (2.6,8);
	\coordinate (A11) at (3,0);
	\coordinate (A12) at (3,8);
	\coordinate (A13) at (3.4,0);
	\coordinate (A14) at (3.4,8);
	\coordinate (A15) at (3.7,0);
	\coordinate (A16) at (3.7,8);
	
 	\coordinate (B11) at (6.3,0);
	\coordinate (B12) at (6.3,8);

	\coordinate (D1) at (6.7,0);
	\coordinate (D2) at (6.7,8);
  	\coordinate (D3) at (7.1,0);
	\coordinate (D4) at (7.1,8);
 
 	\coordinate (E1) at (7.3,0);
	\coordinate (E2) at (7.3,8);
  	\coordinate (E3) at (7.7,0);
	\coordinate (E4) at (7.7,8);

	 
	\draw[very thin] (0.4,4) -- (3.7,4) -- (3.7,6) -- (0.4,6) -- cycle;
	\fill[gray!10] (0.4,4) -- (3.7,4) -- (3.7,6) -- (0.4,6) -- cycle;

	\draw[very thick] (A1)--(A4);
	\draw[very thick] (A2)--(A5);
	\draw[very thick] (A3)--(A6);
	\draw[very thick] (2.4,4)--(2.4,6);
	\draw[very thick] (2.6,4)--(2.6,6);
	\draw[very thick] (3,4)--(3,6);
	\draw[very thick] (3.4,4)--(3.4,6);
	\draw[very thick] (3.7,4)--(3.7,6);

	\fill[gray!30] (A1) -- (A2) -- (A5) -- (A4) -- (A1)  -- cycle;
	\fill[gray!30] (A3) -- (2.4,4) -- (2.4,6) -- (A6) -- (A3)  -- cycle;
	\fill[gray!30] (2.6,4) -- (3,4) -- (3,6) -- (2.6,6) -- (2.6,4)  -- cycle;
	\fill[gray!30] (3.4,4) -- (3.7,4) -- (3.7,6) -- (3.4,6) -- (3.4,4)  -- cycle;
	
	
	\begin{scope}[shift={(0,-.2)}]
	\draw[very thin] (2.6,0) -- (7.7,0) -- (7.7,2) -- (2.6,2) -- (2.6,0) -- cycle;
	\fill[gray!10] (2.6,0) -- (7.7,0) -- (7.7,2) -- (2.6,2) -- (2.6,0) -- cycle;
	
	\coordinate (B1) at (4.3,0);
	\coordinate (B2) at (4.3,2);
  	\coordinate (B3) at (4.6,0);
	\coordinate (B4) at (4.6,2);
	\coordinate (B5) at (4.9,0);
	\coordinate (B6) at (4.9,2);
 	\coordinate (B7) at (5.3,0);
	\coordinate (B8) at (5.3,2);
	\coordinate (B9) at (5.7,0);
	\coordinate (B10) at (5.7,2);

	\draw[very thick] (2.6,0)--(2.6,2);
	\draw[very thick] (3,0)--(3,2);
	\draw[very thick] (3.4,0)--(3.4,2);
	\draw[very thick] (3.7,0)--(3.7,2);
	\draw[very thick] (4.3,0)--(4.3,2);
	\draw[very thick] (4.6,0)--(4.6,2);
	\draw[very thick] (B5)--(B6);
	\draw[very thick] (B7)--(B8);
	\draw[very thick] (B9)--(B10);
	\draw[very thick] (6.3,0)--(6.3,2);
	\draw[very thick] (6.7,0)--(6.7,2);
	\draw[very thick] (7.1,0)--(7.1,2);
	\draw[very thick] (7.3,0)--(7.3,2);
	\draw[very thick] (7.7,0)--(7.7,2);

	\fill[gray!30] (2.6,0) -- (3,0) -- (3,2) -- (2.6,2) -- (2.6,0)  -- cycle;
	\fill[gray!30] (3.4,0) -- (3.7,0) -- (3.7,2) -- (3.4,2) -- (3.4,0)  -- cycle;
	\fill[gray!30] (B1) -- (B3) -- (B4) -- (B2) -- (B1)  -- cycle;
         \fill[gray!30] (B5) -- (B7) -- (B8) -- (B6) -- (B5)  -- cycle;
         \fill[gray!30] (5.7,0) -- (6.3,0) -- (6.3,2) -- (5.7,2) -- (5.7,0)  -- cycle;
	\fill[gray!30] (6.7,0) -- (7.1,0) -- (7.1,2) -- (6.7,2) -- (6.7,0)  -- cycle;
	\fill[gray!30] (7.3,0) -- (7.7,0) -- (7.7,2) -- (7.3,2) -- (7.3,0)  -- cycle;
\end{scope}

	\begin{scope}[shift={(0,.1)}]

	\draw[very thin] (2.6,6) -- (7.7,6) -- (7.7,8) -- (2.6,8) -- (2.6,6) -- cycle; 
	\fill[gray!10] (2.6,6) -- (7.7,6) -- (7.7,8) -- (2.6,8) -- (2.6,6) -- cycle; 
	
	\coordinate (C1) at (4.5,6);
	\coordinate (C2) at (4.5,8);
  	\coordinate (C3) at (4.9,6);
	\coordinate (C4) at (4.9,8);
 	\coordinate (C5) at (5.5,6);
	\coordinate (C6) at (5.5,8);
	
	\draw[very thick] (2.6,6)--(2.6,8);
	\draw[very thick] (3,6)--(3,8);
	\draw[very thick] (3.4,6)--(3.4,8);
	\draw[very thick] (3.7,6)--(3.7,8);
        	\draw[very thick] (6.3,6)--(6.3,8);
	\draw[very thick] (C1)--(C2);
	\draw[very thick] (C3)--(C4);
	\draw[very thick] (C5)--(C6);
	\draw[very thick] (6.7,6)--(6.7,8);
	\draw[very thick] (7.1,6)--(7.1,8);
	\draw[very thick] (7.3,6)--(7.3,8);
	\draw[very thick] (7.7,6)--(7.7,8);

	\fill[gray!30] (2.6,6) -- (3,6) -- (3,8) -- (2.6,8) -- (2.6,6)  -- cycle;
	\fill[gray!30] (3.4,6) -- (3.7,6) -- (3.7,8) -- (3.4,8) -- (3.4,6)  -- cycle;
	\fill[gray!30] (C1) -- (C3) -- (C4) -- (C2) -- (C1)  -- cycle;
         \fill[gray!30] (5.5,6) -- (6.3,6) -- (6.3,8) -- (5.5,8) -- (5.5,6)  -- cycle;
         	\fill[gray!30] (6.7,6) -- (7.1,6) -- (7.1,8) -- (6.7,8) -- (6.7,6)  -- cycle;
	\fill[gray!30] (7.3,6) -- (7.7,6) -- (7.7,8) -- (7.3,8) -- (7.3,6)  -- cycle;
	
	\end{scope}

	\draw[very thin] (6.7,4) -- (7.7,4) -- (7.7,6) -- (6.7,6) -- (6.7,4) -- cycle; 
	\fill[gray!10] (6.7,4) -- (7.7,4) -- (7.7,6) -- (6.7,6) -- (6.7,4) -- cycle; 
	
	\draw[very thick] (6.7,6)--(6.7,4);
	\draw[very thick] (7.1,6)--(7.1,4);
	\draw[very thick] (7.3,6)--(7.3,4);
	\draw[very thick] (7.7,6)--(7.7,4);

         	\fill[gray!30] (6.7,4) -- (7.1,4) -- (7.1,6) -- (6.7,6) -- (6.7,4)  -- cycle;
	\fill[gray!30] (7.3,4) -- (7.7,4) -- (7.7,6) -- (7.3,6) -- (7.3,4)  -- cycle;

	\begin{scope}[shift={(0,-.1)}]

	\draw[very thin] (2.6,2) -- (3.7,2) -- (3.7,4) -- (2.6,4) -- (2.6,2) -- cycle; 
	\fill[gray!10] (2.6,2) -- (3.7,2) -- (3.7,4) -- (2.6,4) -- (2.6,2) -- cycle; 
	
	\draw[very thick] (2.6,2)--(2.6,4);
	\draw[very thick] (3,2)--(3,4);
	\draw[very thick] (3.4,2)--(3.4,4);
	\draw[very thick] (3.7,2)--(3.7,4);

         	\fill[gray!30] (2.6,2) -- (3,2) -- (3,4) -- (2.6,4) -- (2.6,2)  -- cycle;
	\fill[gray!30] (3.4,2) -- (3.7,2) -- (3.7,4) -- (3.4,4) -- (3.4,2)  -- cycle;
\end{scope}
	\begin{scope}[shift={(0,-.1)}]

	\draw[very thin] (6.7,2) -- (7.7,2) -- (7.7,4) -- (6.7,4) -- (6.7,2) -- cycle; 
	\fill[gray!10] (6.7,2) -- (7.7,2) -- (7.7,4) -- (6.7,4) -- (6.7,2) --  cycle; 
	
	\draw[very thick] (6.7,2)--(6.7,4);
	\draw[very thick] (7.1,2)--(7.1,4);
	\draw[very thick] (7.3,2)--(7.3,4);
	\draw[very thick] (7.7,2)--(7.7,4);

         	\fill[gray!30] (6.7,2) -- (7.1,2) -- (7.1,4) -- (6.7,4) -- (6.7,2)  -- cycle;
	\fill[gray!30] (7.3,2) -- (7.7,2) -- (7.7,4) -- (7.3,4) -- (7.3,2)  -- cycle;
		
\end{scope}


	\draw (2.83,7.33) node {$h_1$};
	\draw (3.225,6.66) node {$w_1$};
	\draw (3.57,7.33) node {$h_2$};
	\draw (4.1,6.66) node {$w_2$};
	\draw (4.72,7.33) node {$h_3$};
	\draw (5.2,6.66) node {$w_3$};
	\draw (5.92,7.33) node {$h_4$};
	\draw (6.51,6.66) node {$w_4$};
	\draw (6.9,7.33) node {$h_5$};
	\draw (7.25,6.66) node {$w_5$};
	\draw (7.51,7.33) node {$h_6$};


		\end{tikzpicture}
		\caption{The six slices  corresponding to the deformed good region of Fig.~\ref{fig3}.
For ease of visualization, the slices are drawn slightly detached from each other.
As an example, in the first slice we attached the labels $h_i$ and $w_i$ indicating the widths of the stripes (dark grey regions) and of the rectangular regions separating two subsequent stripes
from each other (light grey regions).}\label{fig4}
\end{figure}

The deformed region $G'$ obtained in the previous step is a union of connected horizontal slices $g_j$ 
of height $\ell$, with $j=1,\ldots, N_g$, 
as shown in Fig.~\ref{fig4}. To each slice we associate a sequence of integers $(h_1,w_1,\ldots,w_{n-1},h_n)$, where $n$ 
is the number of bubbles in $\mathcal B_{G'}$ intersecting the slice, $h_1,h_2,...,h_n$ are their widths, ordered from left to 
right, and $w_1,w_2, ...$ are the spacings between the first and second bubbles, second and third, etc. 

\begin{figure}
\centering
\begin{tikzpicture}
	\coordinate (P1) at (2,0); 
	\coordinate (P2) at (8,0); 
	\coordinate (P3) at (8,8); 
	\coordinate (P4) at (2,8); 
	\coordinate (P5) at (2,6);
	\coordinate (P6) at (0,6);  
	\coordinate (P7) at (0,4);
	\coordinate (P8) at (2,4);
	\coordinate (P9) at (4,2);
	\coordinate (P10) at (6,2);    
	\coordinate (P11) at (6,6);    
	\coordinate (P12) at (4,6);    
	
	\draw[very thin] (A9) -- (E3) -- (E4) -- (2.6,8) -- (2.6,6) -- (A4) -- (A1) -- (2.6,4) -- (A9) -- cycle;

	\fill[gray!10] (A9) -- (E3) -- (E4) -- (2.6,8) -- (2.6,6) -- (A4) -- (A1) -- (2.6,4) -- (A9) -- cycle;
	
	\draw[very thin] (3.7,2) -- (6.7,2) -- (6.7,6) -- (3.7,6) -- (3.7,2)  -- cycle;

	\fill[white] (3.7,2) -- (6.7,2) -- (6.7,6) -- (3.7,6) -- (3.7,2)  -- cycle;
	 
	\coordinate (A1) at (0.4,4);
	\coordinate (A2) at (0.9,4);
	\coordinate (A3) at (1.7,4);
	\coordinate (A4) at (0.4,6);
	\coordinate (A5) at (0.9,6);
	\coordinate (A6) at (1.7,6);
         \coordinate (A7) at (2.4,0);
	\coordinate (A8) at (2.4,8);
  \coordinate (A9) at (2.6,0);
	\coordinate (A10) at (2.6,8);
 \coordinate (A11) at (3,0);
	\coordinate (A12) at (3,8);
 \coordinate (A13) at (3.4,0);
	\coordinate (A14) at (3.4,8);
 \coordinate (A15) at (3.7,0);
	\coordinate (A16) at (3.7,8);

	\draw[very thick] (A1)--(A4);
	\draw[very thick] (A2)--(A5);
	\draw[very thick] (A3)--(A6);
	\draw[very thick] (2.4,4)--(2.4,6);

	\fill[gray!30] (A1) -- (A2) -- (A5) -- (A4) -- (A1)  -- cycle;
	
	\fill[gray!30] (A3) -- (2.4,4) -- (2.4,6) -- (A6) -- (A3)  -- cycle;
	
	\draw[very thick] (A9)--(A10);
	\draw[very thick] (A11)--(A12);
	\draw[very thick] (A13)--(A14);
\draw[very thick] (A15)--(A16);

	\fill[gray!30] (A9) -- (A11) -- (A12) -- (A10) -- (A9)  -- cycle;
	\fill[gray!30] (A13) -- (A15) -- (A16) -- (A14) -- (A13)  -- cycle;
	
	   \coordinate (B1) at (4.3,0);
	\coordinate (B2) at (4.3,2);
  \coordinate (B3) at (4.6,0);
	\coordinate (B4) at (4.6,2);
 \coordinate (B5) at (4.9,0);
	\coordinate (B6) at (4.9,2);
 \coordinate (B7) at (5.3,0);
	\coordinate (B8) at (5.3,2);
 \coordinate (B9) at (5.7,0);
	\coordinate (B10) at (5.7,2);
 \coordinate (B11) at (6.3,0);
	\coordinate (B12) at (6.3,8);

	\draw[very thick] (B1)--(B2);
	\draw[very thick] (B3)--(B4);
	\draw[very thick] (B5)--(B6);
\draw[very thick] (B7)--(B8);
\draw[very thick] (B9)--(B10);
\draw[very thick] (6.3,0)--(6.3,2);
\draw[very thick] (6.3,6)--(6.3,8);

	  \coordinate (C1) at (4.5,6);
	\coordinate (C2) at (4.5,8);
  \coordinate (C3) at (4.9,6);
	\coordinate (C4) at (4.9,8);
 \coordinate (C5) at (5.5,6);
	\coordinate (C6) at (5.5,8);

	\draw[very thick] (C1)--(C2);
	\draw[very thick] (C3)--(C4);
	\draw[very thick] (C5)--(C6);

	\fill[gray!30] (C1) -- (C3) -- (C4) -- (C2) -- (C1)  -- cycle;
	\fill[gray!30] (B1) -- (B3) -- (B4) -- (B2) -- (B1)  -- cycle;
         \fill[gray!30] (B5) -- (B7) -- (B8) -- (B6) -- (B5)  -- cycle;
         \fill[gray!30] (5.7,0) -- (6.3,0) -- (6.3,2) -- (5.7,2) -- (5.7,0)  -- cycle;
         \fill[gray!30] (5.5,6) -- (6.3,6) -- (6.3,8) -- (5.5,8) -- (5.5,6)  -- cycle;
         
	  \coordinate (D1) at (6.7,0);
	\coordinate (D2) at (6.7,8);
  \coordinate (D3) at (7.1,0);
	\coordinate (D4) at (7.1,8);
 
 	  \coordinate (E1) at (7.3,0);
	\coordinate (E2) at (7.3,8);
  \coordinate (E3) at (7.7,0);
	\coordinate (E4) at (7.7,8);
 
	\draw[very thick] (D1)--(D2);
	\draw[very thick] (D3)--(D4);
	\draw[very thick] (E1)--(E2);
	\draw[very thick] (E3)--(E4);

	\fill[gray!30] (D1) -- (D3) -- (D4) -- (D2) -- (D1)  -- cycle;
	\fill[gray!30] (E1) -- (E3) -- (E4) -- (E2) -- (E1)  -- cycle;

\draw[red, very thick] (0.4,4) -- (0.9,4); \draw (0.7,3.8) node {$s_{17}$};
\draw[red, very thick] (0.4,6) -- (0.9,6); \draw (0.7,6.2) node {$s_1$};
\draw[red, very thick] (1.7,4) -- (2.4,4); \draw (2.1,3.8) node {$s_{16}$};
\draw[red, very thick] (1.7,6) -- (2.4,6); \draw (2.1,6.2) node {$s_2$};
\draw[red, very thick] (2.6,0) -- (3,0); \draw (2.8,-0.2) node {$s_{15}$};
\draw[red, very thick] (2.6,8) -- (3,8); \draw (2.8,8.2) node {$s_3$};
\draw[red, very thick] (3.4,0) -- (3.7,0); \draw (3.6,-0.2) node {$s_{14}$};
\draw[red, very thick] (3.4,8) -- (3.7,8); \draw (3.6,8.2) node {$s_4$};
\draw[red, very thick] (4.3,0) -- (4.6,0); \draw (4.45,-0.2) node {$s_{13}$};
\draw[red, very thick] (4.3,2) -- (4.6,2); \draw (4.45,2.2) node {$s_{18}$};
\draw[red, very thick] (4.9,0) -- (5.3,0); \draw (5.1,-0.2) node {$s_{12}$};
\draw[red, very thick] (4.9,2) -- (5.3,2); \draw (5.1,2.2) node {$s_{19}$};
\draw[red, very thick] (5.7,0) -- (6.3,0); \draw (6,-0.2) node {$s_{11}$};
\draw[red, very thick] (5.7,2) -- (6.3,2); \draw (6,2.2) node {$s_{20}$};
\draw[red, very thick] (6.7,0) -- (7.1,0); \draw (6.9,-0.2) node {$s_{10}$};
\draw[red, very thick] (6.7,8) -- (7.1,8); \draw (6.9,8.2) node {$s_7$};
\draw[red, very thick] (7.3,0) -- (7.7,0); \draw (7.5,-0.2) node {$s_9$};
\draw[red, very thick] (7.3,8) -- (7.7,8); \draw (7.5,8.2) node {$s_8$};
\draw[red, very thick] (4.5,6) -- (4.9,6); \draw (4.7,5.8) node {$s_{22}$};
\draw[red, very thick] (4.5,8) -- (4.9,8); \draw (4.7,8.2) node {$s_5$};
\draw[red, very thick] (5.5,6) -- (6.3,6); \draw (5.9,5.8) node {$s_{21}$};
\draw[red, very thick] (5.5,8) -- (6.3,8); \draw (5.9,8.2) node {$s_6$};

		\end{tikzpicture}
\caption{The red segments $s_j$ (color online) for the bubble configuration of Fig.~\ref{fig3}
in the deformed good region $G'$.
The segments can be naturally grouped in pairs: two segments form a pair if they belong to the boundary of the same droplet. E.g., $s_1$ and $s_{17}$ are paired, $s_{11}$ and $s_{20}$ are paired, 
etc. Every segment $s_j$ comes with its length $h(s_j)$, and with two spacings, the left spacing $w_1(s_j)$ and the right spacing $w_2(s_j)$, corresponding to the distances to the closest droplets 
to its left and to its right, by moving along the boundary. If one of the endpoints of $s_j$ is a corner of 
the boundary of $G'$, then $s_j$ may not have any droplet to its left or right, in which case we let
the corresponding spacing to be infinite. In the example in the figure, $w_1(s_1)=w_1(s_{17})=
w_2(s_8)=w_2(s_9)=+\infty$, and all the other spacings are finite. }\label{fig5}\end{figure}
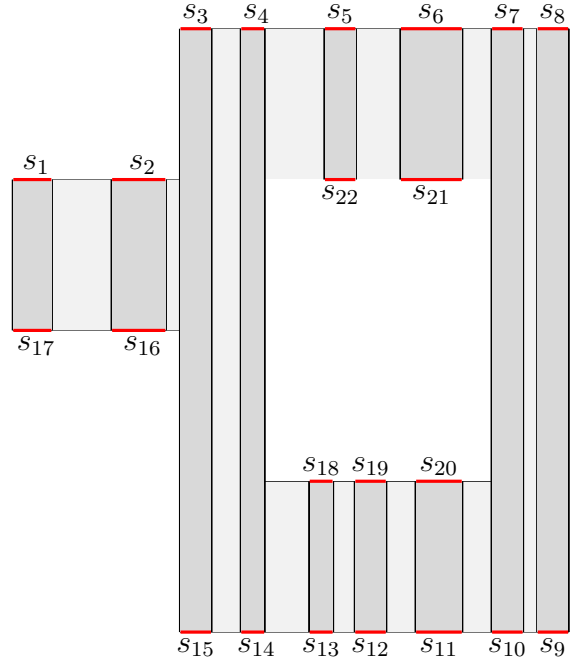

We also denote by $s_i$, with $i=1,\ldots,N_s$, the maximal connected segments in the intersection between the horizontal 
boundary of $G'$ and the boundaries of the droplets in $\mathcal B_{G'}$ (note that the boundary of a 
rectangular bubble does not coincide with its contour: rather, it consists of four segments, two horizontal 
and two vertical). Note that segments come in pairs: one can say that two segments form a pair if they belong to the boundary of the same droplet. 
Moreover, to each segment $s_j$ we associate its length $h(s_j)$ and two spacings $w_1(s_j)$, 
$w_2(s_j)$, which are the horizontal distances from the next droplets (``next'' by following the boundary of $G'$) to the left and to the right of $s_j$. Note that if $s_j$ touches a corner 
of $\partial G'$,  say on its right side, then there may not be any droplet to its right: by following 
the boundary one may find that the next segment to the right could actually have the same 
horizontal coordinates, in which case we will assign the value $+\infty$ to $w_2(s_j)$ (and similarly for $w_1(s_j)$ in the case of the next segment on the left); see Fig.~\ref{fig5}.

The key ingredient for the proof of Lemma \ref{lm2.2} is the following.

\begin{lemma}\label{lm2.3}
Given $G'$ and $\mathcal B_{G'}$ as above, we have: 
\be E_{G'}(\mathcal B_{G'})\ge \ell\sum_{j=1}^{N_g} e_{\infty}(h_1^{(j)},w_1^{(j)},\ldots, h_{n_j}^{(j)})-
\sum_{j=1}^{N_s} f(w_1(s_j),h(s_j),w_2(s_j)),\label{2.13}\ee
where $h_1^{(j)},w_1^{(j)},\ldots, h_{n_j}^{(j)}$ is the sequence of widths and spacings associated to the slice $g_j$, and the functions $e_\infty$ and $f$ are defined as follows: 
\bea  e_{\infty}(h_1,w_1,\ldots,h_n)&=&4Jn-2\sum_{i=1}^n\sum_{\xx\in\mathbb Z^2\setminus\{\bf 0\}}
\frac{\min\{|x_1|,h_i\}}{|\xx|^p}  +\nonumber\\
& &+\frac12 \sum_{\substack{i,j=1,\ldots,n\\ i\neq j}} W(l_i,\mathcal L(l_j))\,,\label{2.14bis}\eea
where $l_i=\{(x,0)\in \mathbb Z^2:\  0<x-\sum_{k=1}^{i-1}(h_k+w_k)\le h_i\}$
and $\mathcal L(l_i)$ is the smallest infinite vertical strip containing $l_i$, that is
$\mathcal L(l_i)=\{(x,y)\in \mathbb Z^2:\  0<x-\sum_{k=1}^{i-1}(h_k+w_k)\le h_i,\ y\in\mathbb Z\}$; moreover,  
\be f(w_1,h,w_2)=\frac12W(\mathcal L^+_h,Q_{w_1,h,w_2})\ee
where $\mathcal L^+_h=\{(x,y)\in \mathbb Z^2:\  -h\le x< 0, y>0\}$ and $Q_{w_1,h,w_2}=\{(x,y)\in \mathbb Z^2:\  x<-w_1-h\ {\rm or}\ x\ge w_2,\  y\le 0\}$.
\end{lemma}

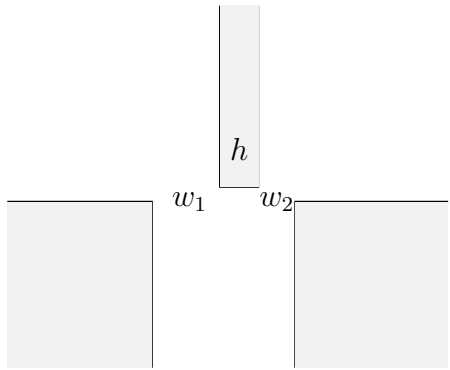
\begin{figure}
\centering
\begin{tikzpicture}
	
	\draw[thick] (2,2.4) -- (2,0) -- (2.5,0) -- (2.5,2.4);
	
	\fill[gray!10] (2,2.4) -- (2,0) -- (2.5,0) -- (2.5,2.4) -- (2.5,2.4) -- cycle;
	
	\draw[thick] (2.99,-2.4) -- (2.99,-0.2) -- (5,-0.2);
	
	\fill[gray!10] (2.99,-2.4) -- (2.99,-0.2) -- (5,-0.2) --(5,-2.4) -- (2.99,-2.4) -- cycle;

	\draw[thick]  (-.8,-0.2) -- (1.1,-0.2) -- (1.1,-2.4);
	
	\fill[gray!10]  (-.8,-0.2) -- (1.1,-0.2) -- (1.1,-2.4) -- (-.8,-2.4) -- (-.8,-.2) -- cycle;
\draw (1.6,-0.2) node {$w_1$};
\draw (2.75,-0.2) node {$w_2$};
\draw (2.25,0.5) node {$h$};

\end{tikzpicture}
\caption{A pictorial representation of the droplets involved in the definitions of $f(w_1,h,w_2)$,
which represents the interaction energy of the infinite half strip of width $h$, with the two quarter spaces to its lower left and lower right. The spacings $w_1$ and $w_2$ represent the horizontal distances between the infinite half strip and the quarter space to its left and to its right, respectively. }\label{fig6}\end{figure}

{\bf Remark.} $e_\infty$ is the energy per unit vertical length of an infinite vertically striped configuration,
and $f(w_1,h,w_2)$ is the interaction energy between the droplets in Fig.~\ref{fig6}.

\medskip 

{\bf Proof of Lemma \ref{lm2.3}.} First of all, note that the contribution 
$$\sum_{\b\in\mathcal B_{G'}}[2J|\G_\b|+u_{G'}(\b)]$$ to the energy $E_{G'}(\mathcal B_{G'})$ (see \eqref{eq2.8}) is identical to the 
corresponding contribution in $\ell\sum_{j=1}^{N_g} e_{\infty}(h_1^{(j)},w_1^{(j)},\ldots, h_{n_j}^{(j)})$,
i.e., to the one arising from the first two terms on the right side of \eqref{2.14bis}.
Therefore, all we have to prove is that the difference between the interaction terms in $E_{G'}(\mathcal B_{G'})$ and in $\ell\sum_{j=1}^{N_g} e_{\infty}(h_1^{(j)},w_1^{(j)},\ldots, h_{n_j}^{(j)})$ is bounded from below by the last sum in \eqref{2.13}.

Let us focus on a given slice $g_j$ and on the intersection of a given bubble $\b$ 
with this slice. The interaction energy of this portion of bubble with all the other bubbles in $G'$,
as it appears in $E_{G'}(\mathcal B_{G'})$, is 
\be \frac12 \sum_{\b'\in\mathcal B_{G'}}^* W(\d_\b\cap g_j,\d_{\b'})\,,\label{2.14}\ee
where we recall that the $*$ on the sum indicates the constraint that the bubbles $\b'$ intersecting $\b$
after vertical translations should {\it not} be included in the sum. Eq.~(\ref{2.14}) can be bounded from 
below by summing only over the bubbles $\b'$ having non zero intersection with $g_j$: 
$$(\ref{2.14})\ge \frac12 \sum_{\substack{\b'\in\mathcal B_{G'}\\
\b'\cap g_j\neq\emptyset,\ \b'\neq\b}} W(\d_\b\cap g_j,\d_{\b'}).$$
The term $\ell\sum_{j=1}^{N_g} e_{\infty}(h_1^{(j)},w_1^{(j)},\ldots, h_{n_j}^{(j)})$ on the right side of 
\eqref{2.13} contains a term of that form, with the difference that 
$\d_{\b'}$ is replaced by the infinite vertical strip of the same width containing it. In fact, the interaction 
term in \eqref{2.14bis} satisfies
\be \ell\sum_{j=1}^{N_g} \sum_{\substack{i,i'=1,\ldots,n_j\\ i\neq i'}} W(l_i^{(j)},\mathcal L(l^{(j)}_{i'}))=
\sum_{j=1}^{N_g}\sum_{\b\in\mathcal B_{G'}}
\sum_{\substack{\b'\in\mathcal B_{G'}\\
\b'\cap g_j\neq\emptyset,\ \b'\neq\b}} W(\d_\b\cap g_j,\mathcal L(\d_{\b'})),
\ee
where $l_i^{(j)}$ is the analogue of the set $l_i$ defined after \eqref{2.14bis}, corresponding to the 
widths and spacings $h_1^{(j)},w_1^{(j)},\ldots, h_{n_j}^{(j)}$ associated to the slice $g_j$.
Therefore, what remains to be proved is that 
\be \label{desi} \sum_{j=1}^{N_g}\sum_{\b\in\mathcal B_{G'}}
\sum_{\substack{\b'\in\mathcal B_{G'}\\
\b'\cap g_j\neq\emptyset,\ \b'\neq\b}} W(\d_\b\cap g_j,\mathcal L(\d_{\b'})\setminus \d_{\b'})
\le \sum_{j=1}^{N_s} W(\mathcal L^+_{h(s_j)},Q_{w_1(s_j), h(s_j), w_2(s_j)}).
\ee

Let us rewrite the left side of \eqref{desi} as 
\be  \sum_{\b'\in\mathcal B_{G'}} \sum_{\substack{j=1,...,N_g\\
g_j\cap \b'\neq\emptyset}} \sum_{\substack{\b\in\mathcal B_{G'}\\ \b\neq\b'}}
W(\mathcal L(\d_{\b'})\setminus \d_{\b'},\d_\b\cap g_j).\ee
Note that the horizontal boundary of $\mathcal L(\d_{\b'})\setminus \d_{\b'}$ consists of two segments 
$s_j, s_{k}$ with $h(s_j)=h(s_k)$ and $s_j$ above $s_k$ (these are the pairs of segments mentioned 
in Section \ref{sec2.4.2}; see Fig.~\ref{fig5}), and there is a one-to-one correspondence 
between summing over $\b'$ and summing over 
these pairs of segments $s_j,s_k$ (which is of course the same as summing over all segments).
Moreover, the set $\mathcal L(\d_{\b'})\setminus \d_{\b'}$ is equal to the union of two sets, each of which 
is a translation, and one also a reflection, of 
$\mathcal L^+_{h(s_j)}$, that is
\be \mathcal L(\d_{\b'})\setminus \d_{\b'}=\t^1_{\b'}\mathcal L^+_{h(s_j)}\cup \t^2_{\b'}r
\mathcal L^+_{h(s_j)}\ee
for translations $\t^1_{\b'}$ and $\t^2_{\b'}$, and $r$ denoting reflection about the $x_1$-axis. 

To conclude, it is enough to note that 
\be \bigcup_{\substack{i=1,...,N_g\\
g_i\cap \b'\neq\emptyset}} \bigcup_{\b\in\mathcal B_{G'}} \d_\b\cap g_i
\subseteq \t^1_{\b'}Q_{w_1(s_j),h(s_j),w_2(s_j)}
\cap \t^2_{\b'}rQ_{w_1(s_k),h(s_k),w_2(s_k)},\ee
which implies the desired inequality  \eqref{desi}, and thus completes the proof of Lemma~\ref{lm2.3}.\qed

\subsubsection{Reflection positivity}

We now show how to bound from below the right side of \eqref{2.13}, and how to use the resulting estimate to conclude the proof of Lemma \ref{lm2.2}. The key step is to use  reflection positivity to obtain a lower bound on the $e_\infty$
term. This is an application of the block reflection positivity for one-dimensional spin systems worked out in \cite{GLL06}. The result is the following: 
\be e_\infty(h_1,w_1,\ldots,h_n)\ge \t+\sum_{i=1}^n h_ie_{\rm s}(h_i)+
\sum_{i=1}^{n-1}w_i e_{\rm s}(w_i).\label{2.23}\ee
To see this, note that $e_\infty(h_1,w_1,\ldots,h_n)=\lim_{L\to\infty}H^{\rm per} _{\L_L}(\ss_{h_1,\ldots, h_n})$, where: 
\begin{enumerate}
\item  given a spin configuration $\ss_{\L_L}$ on $\L_L=[1,L]\cap \mathbb Z$, 
\be H^{\rm per} _{\L_L}(\ss_{\L_L})=-J\sum_{i=1}^L(\s_i\s_{i+1}-1)+\sum_{1\le i<j\le L}
(\s_i\s_j-1)v_L(i-j)\,,\ee
with \be v_L(x)=\sum_{n,y\in\mathbb Z}\big((x+nL)^2+y^2\big)^{-p/2},\label{vl}\ee
and $\sigma_{L+1}\equiv \sigma_{1}$;
\item the spins in the  configuration $\ss_{h_1,\ldots, h_n}$ are equal to $-1$ on 
the intervals $\{x\in \mathbb Z:\  0<x-\sum_{k=1}^{i-1}(h_k+w_k)\le h_i\}$, and $+1$ otherwise.
\end{enumerate}
Now, $H^{\rm per} _{\L_L}(\ss_{h_1,\ldots, h_n})$ is a one-dimensional spin Hamiltonian with a reflection positive long-range interaction
and periodic boundary conditions, of the class considered in \cite{GLL06,GLL07}.
Therefore, we can apply the chessboard estimate 
proved e.g. in the Appendix 
of \cite{GLL07}. As a result, using \cite[Eqs.~(A4)--(A5)]{GLL07} and recalling the fact that the spin configuration 
$\ss_{h_1,\ldots, h_n}$ consists 
of blocks of alternating sign, of size $h_1,w_1,\ldots, h_n,w_{n}$, with $w_n=w_n(L)=L-(h_1+w_1+\cdots+h_n)$, we get
\be H^{\rm per} _{\L_{L}}(\ss_{\L_{L}})\ge \sum_{i=1}^{n}(h_i e_{\rm s}(h_i)+w_i e_{\rm s}(w_i))\;,\label{C.3}\ee
where $e_{\rm s}(h)$ is the energy per site (as computed from $H^{\rm per} _{\L_{L}}$, in the limit 
$L\to\infty$) of the infinite periodic configuration consisting of blocks all of the same size 
$h$, and  of alternating sign. Note that $e_{\rm s}(h)$ is the same as the one defined in Section \ref{sec1} for the two-dimensional model. Finally, to go from \eqref{C.3} to \eqref{2.23}, observe that $\lim_{L\to\infty}w_n(L)e_{\rm s}(w_n(L))=\t$. This follows, e.g., from the explicit expression of $e_{\rm s}(h)$, derived in \cite[Appendix A]{GLS}: 
\be e_{\rm s}(h)=\frac{\tau}{h}+\frac2{h}\int_0^\infty\,d\alpha\,\mu_v(\alpha)\frac{e^{-\alpha}}{(1-e^{-\alpha})^2}(1-\tanh\frac{\alpha h}{2})=\frac{\tau}{h}+\frac{A_p}{h^{p-2}}+O(h^{-p})
\label{app0.4}\ee
for large $h$, where $\mu_v(\a)$ is the 
inverse Laplace transform of the function $v_\infty(x)$ in \eqref{vl}, i.e., the function such that $v_{\infty}(x)=\int_0^\infty d\a\,\m_v(\a) e^{-\a x}$, 
$\forall x>0$, and $A_p$ is a suitable constant.

\medskip

{\bf Remark.} From Eq.~(\ref{app0.4}) it follows straight away that the optimal stripe width is $h^* = ((p-2)A_p |\t|^{-1})^{1/(p-3)} (1+o(1))$ as $\tau\to 0$, and also that $e_s(h^*) = \tfrac{p-3}{p-2} \tfrac{\tau}{h^*}(1+ o(1))$.

\subsubsection{Putting things together}

Plugging \eqref{2.23} into \eqref{2.13} gives
\bea && E_{G'}(\mathcal B_{G'})\ge |G'|e_{\rm s}(h^*)+\ell\t N_g\nonumber\\
&&+\ell 
\sum_{j=1}^{N_g} \Big[\sum_{k=1}^{n_j}h_k^{(j)} (e_{\rm s}(h_k^{(j)})-e_{\rm s}(h^*))+
\sum_{k=1}^{n_j-1}w_k^{(j)} (e_{\rm s}(w_k^{(j)})-e_{\rm s}(h^*))\Big]\nonumber
\\
&&-
\sum_{j=1}^{N_s} f(w_1(s_j),h(s_j),w_2(s_j))\,.\label{2.26}\eea
Now, $\ell N_g\le \tfrac12|\partial G'|$ and it remains to show that the sum of the last two lines can be bounded from below by (const.)$\t|\partial G'|+\frac12
\sum_{h\neq h^*} (e_{\rm s}(h)-e_{\rm s}(h^*))A_h(G)$. From the definition of $f(w_1,h,w_2)$ it easily follows that it can be bounded independently of $h$, as
\be f(w_1,h,w_2)\le \sum_{i=1,2}\frac{C_2}{w_i^{p-4}}\label{2.29}\ee
for a suitable constant $C_2$. Moreover, from \eqref{app0.4} it follows that 
\be e_{\rm s}(w)-e_{\rm s}(h^*)\ge \frac{C_3}{w^{p-2}}+\frac{\t}{w}\label{2.30}\ee
for all $w\ge 1$ and a suitable $C_3>0$. Note that the left side is non-negative, while the right side may be negative. A simple consequence of \eqref{2.30} is that 
\be \frac{C_3}{w^{p-4}}\le |\t| w+(C_3|\t|^{-1})^{\frac1{p-3}}w\big(e_{\rm s}(w)-e_{\rm s}(h^*)\big). \label{2.34x}
\ee
Using this bound in \eqref{2.29} implies that the last line of \eqref{2.26} can be bounded as
\bea && \sum_{j=1}^{N_s} f(w_1(s_j),h(s_j),w_2(s_j))\le\label{2.31}\\
&&\le  \frac{C_2}{C_3}\sum_{j=1}^{N_s}\sideset{}{'}
\sum_{i=1,2}\Big[w_i(s_j)|\t|+(C_3|\t|^{-1})^{\frac1{p-3}}w_i(s_j)\big(e_{\rm s}(w_i(s_j))-e_{\rm s}(h^*)\big)\Big],\nonumber
\eea
where the prime on the sum indicates the constraint that $w_i(s_j)<\infty$. Now, 
every spacing $w_i(s_j)$ appears twice in the sum above (because every spacing is to the left or to the right of two different segments $s_j, s_{j'}$), hence
$\sum_{j=1}^{N_s}
\sum_{i=1,2}' w_i(s_j)\le 2|\partial G'|$. Similarly, 
\be \sum_{j=1}^{N_s}\sideset{}{'}
\sum_{i=1,2}w_i(s_j)\big(e_{\rm s}(w_i(s_j))-e_{\rm s}(h^*)\big)\le 
2
\sum_{j=1}^{N_g}
\sum_{k=1}^{n_j-1}w_k^{(j)} (e_{\rm s}(w_k^{(j)})-e_{\rm s}(h^*)).\ee
Therefore, if $\ell\ge 4C_2 C_3^{(4-p)/(p-3)}|\t|^{-1/(p-3)}$,
\bea && E_{G'}(\mathcal B_{G'})\ge |G'|e_{\rm s}(h^*)+\t|\partial G'|\Big(\frac12+2\frac{C_2}{C_3}\Big)+
\label{2.34}\\
&&\qquad +
\ell 
\sum_{j=1}^{N_g} \Big[\sum_{k=1}^{n_j}h_k^{(j)} (e_{\rm s}(h_k^{(j)})-e_{\rm s}(h^*))+\frac12
\sum_{k=1}^{n_j-1}w_k^{(j)} (e_{\rm s}(w_k^{(j)})-e_{\rm s}(h^*))\Big].\nonumber\eea
To complete the proof of Lemma \ref{lm2.2}, note that $|G'|\le |G|$, and $|\partial G'|\le 2|\partial G|$,
as already argued above. \qed

\medskip

{\bf Remark.} The proof of \eqref{2.34} is valid for bubble configurations a bit more general than 
those considered here: in fact, we never used the fact that $\mathcal B_{G'}$ has no holes, in the sense 
explained in Section \ref{sec1.1.3}. 
The only property we really used is that the bubbles in $\mathcal B_{G'}$ are all rectangular
with the same orientation.

\subsection{Proof of Lemma \ref{lm1}}\label{app10}

We proceed similarly to \cite[Section 3]{GLS}. The first step is to estimate the cost of erasing the 
bubbles with corners. Write $n_c(T)=\sum_{\b\in \mathcal B_T}\nu_c(\b)$, with $\nu_c(\b)$ the number of corners associated with the bubble $\b$, which may be an integer or a half-integer. 
Consider a bubble with $\nu_c(\b)>0$. Dropping the positive interaction of this bubble with the others, 
its contribution to the energy is bounded from below as
\be  2J|\G_\b|+u_T(\b)+2^{1-p/2}\n_c(\b) \ge
\t |\G_\b|+2^{1-\frac{p}{2}}\n_c(\b)\;.\label{e3.4g}\ee
Note that, in order for $\G_\b$ to be very long, the number of corners must
be sufficiently  large: in formulae (see \cite[Eq.(3.10)]{GLS} and following lines), 
\be |\G_\b|\le 2\ell+2\ell\n_c(\b)\;.\label{d3.10}\ee
If, as we are assuming, $\n_c(\b)\ge 1/2$, then $\n_c(\b)+1\le 3\n_c(\b)$, so that $\n_c(\b)\ge |\G_\b|/(6\ell)$. Inserting this back into (\ref{e3.4g}) gives
\be  2J|\G_\b|+u_T(\b)+2^{1-p/2}\n_c(\b) \ge
2^{-p/2} \frac{|\G_\b|}{6\ell}\Big(1-6\cdot 2^{p/2}|\t|\ell\Big)+2^{-\frac{p}{2}}\n_c(\b)\;.\label{e3.4gg}\ee
The first term on the right side is positive, and, therefore, can be dropped for a lower bound, if 
$\ell< (6\cdot 2^{p/2}|\t|)^{-1}$. Therefore, denoting by $\mathcal S_T$ the subset of $\mathcal B_T$ 
consisting of all the bubbles without corners, 
\be E_T(\mathcal B_T)\ge E_T(\mathcal S_T)+2^{-\frac{p}{2}}n_c(T).\ee

In order to estimate the energy of the corner-less configuration $\mathcal S_T$ we proceed exactly as in 
the proof Lemma \ref{lm2.2}. Assume that the contours in $\mathcal S_T$ are vertical. 
We deform the tile $T$ by moving to the right the left vertical boundary of $T$, until it hits
the left vertical contour of a bubble, and vice versa for the right vertical boundary. We call $T'$ and 
$\mathcal S_{T'}$ the new region and configuration obtained after the deformation. 
In passing from $T,\mathcal S_T$ to $T',\mathcal S_{T'}$ we increase the energy by at most $2|\t|\ell$. 
Now we use the bound \eqref{2.34} which, as remarked after \eqref{2.34}, is valid for 
all configurations consisting only of rectangular bubbles with the same orientation. The result is  
\bea && E_{T'}(\mathcal S_{T'})\ge e_{\rm s}(h^*)|T'|+C\t\ell+\\
&&\qquad +\ell\sum_{i=1}^n h_i\big(e_{\rm s}(h_i)-e_{\rm s}(h^*)\big)+\frac{\ell}2
\sum_{i=1}^{n-1} w_i\big(e_{\rm s}(w_i)-e_{\rm s}(h^*)\big),\nonumber\eea
where $h_1,\ldots,h_n$ are the widths of the bubbles in $\mathcal S_{T'}$, and $w_1,\ldots,w_{n-1}$
their separations.

It $T$ contains a hole, then either one of the 
$h_i$'s or $w_i$'s is larger than $\ell/5$, or, the width of $T'$ is smaller than $4\ell/5$. 
In the first case, one of the  terms $\ell h_i\big(e_{\rm s}(h_i)-e_{\rm s}(h^*)\big)$ or $\ell w_i\big(e_{\rm s}(w_i)-e_{\rm s}(h^*)\big)$ is larger than $(\ell^2/5)|e_{\rm s}(h^*)|$: to see this recall that $\ell\ge c_0 h^*$, for a large enough constant $c_0$, and use \eqref{app0.4}, which implies that $e_s(h)$ is positive for $h\geq \ell/5$ in this case.
In the second case, 
the difference between $e_{\rm s}(h^*)|T'|$ and $e_{\rm s}(h^*)|T|$ is larger than $(\ell^2/5)|e_{\rm s}(h^*)|$. In both cases, we get a gain at least $(\ell^2/5)|e_{\rm s}(h^*)|$, which is larger than (const.)$\ell^2 |\t|^{(p-2)/(p-3)}$. To conclude the proof, note that under the stated assumptions on $\ell$ (that is, $c_0h^*\le \ell\le (c_0|\t|)^{-1}$ for a suitable constant $c_0$), the error term $(2+C)|\t|\ell$ is smaller than $c\big[n_c(T)+|\t|^{(p-2)/(p-3)}\ell^2\c_{\rm hole}(T)
\big]$, where $c$ can be made as small as desired, by increasing $c_0$ 
(recall also that by definition of bad tile, either $n_c(T)\ge 1/2$, or $\c_{\rm hole}(T)=1$).
This concludes the proof. \qed

\appendix
\section{The higher-dimensional case}\label{appa.1}

In this appendix we shall detail our main results in the case $d\geq 3$, and explain the main differences in their proof as compared to the two-dimensional case. The starting point is a representation of the energy in terms of droplets as in \eqref{2.15}, whose boundaries, separating  plus spins from minus spins, consist  now  of $d-1$ dimensional plaquettes. Tiles are now $d$-dimensional cubes of side length $\ell$, and are used to divide space into good regions and bad cubes, with the good regions only containing ``stripes'' (i.e., quasi-one-dimensional regions of uniform spins, delimited by two flat parallel interfaces; they are slabs in $d=3$), which can be oriented in $d$ different directions. 

Our first claim concerns the fact that the localization bound \eqref{2.7} still holds, with the obvious notion of \lq\lq corner\rq\rq, namely $d-2$ dimensional segments where two plaquettes with different orientation meet. The proof of the analogue of \eqref{2.7} in higher dimensions is essentially the same as in $d=2$, and relies on the analogue of \eqref{eq:2.9}, whose proof is in \cite[App.~D]{GLS}. 

The key bound in Lemma~\ref{lm2.2} for the good regions holds verbatim also for general $d\geq 2$.  After the modification from $G$ to $G'$, each stripe has a definite width $h$ but will not be a cuboid, in general; it is bounded by a union of $d-1$ dimensional cuboids $s_k$ with width $h$ and all other dimensions equal to $\ell$. As before, each $s_k$ comes with two numbers,  $w_1(s_k)$ and $w_2(s_k)$, measuring the distance to the next slice in the direction perpendicular to the stripes. The analogues of the slices $g_j$ introduced in Sect.~\ref{sec2.4.2} are cylinders with base area $\ell^{d-1}$ and various heights, which are obtained by adding up the various stripe widths $h_i$ and their separation $w_i$; they are oriented perpendicular to the stripes. With these modified definitions, Eq.~\eqref{2.13} still holds, with $\ell$ replaced by $\ell^{d-1}$ in front of the first term on the right side, and $f(w_1,h,w_2)$ now denoting the interaction energy as depicted in Fig.~\ref{fig6}, with the upper strip of width $h$ extended by $\ell$ in the remaining $d-2$ dimensions, while the two lower ones are infinite in those directions. This function $f$ satisfies the bound
\be\label{new:f}
f(w_1,h,w_2) \leq \ell^{d-2} \sum_{i=1,2} \frac {C_2}{w_i^{p-d-2}}\,.
\ee
As already discussed in \cite[App.~A]{GLS}, the analogue of \eqref{app0.4} for general $d$ is 
\be e_{\rm s}(h)=\frac{\tau}{h}+\frac{A_{p,d}}{h^{p-d}}+O\left(h^{d-p-2}\right)\,,
\label{app0.4app}\ee
from which it follows that $e_{\rm s}(h^*) \sim |\tau|^{(p-d)/(p-d-1)}$ and $h^* \sim |\tau|^{-1/(p-d-1)} $ for small $\t$.
Moreover, one easily deduces that \eqref{new:f} can be bounded by 
\be \frac{C_3}{w^{p-d-2}}\le |\t| w+(C_3|\t|^{-1})^{\frac1{p-d-1}}w\big(e_{\rm s}(w)-e_{\rm s}(h^*)\big)\,, \label{2.34new}
\ee
which is the analogue of Eq.~\eqref{2.34x}. The rest of the proof of Eq.~\eqref{eq:2.8} for general $d$ remains unchanged.

The analogue of Lemma~\ref{lm1} for general $d\geq 2$ takes the following form:

\begin{customthml}{2'}\label{lm1appx}
For given $d\geq 2$, there exist  positive constants $c_0$, $c_2$ and $\e$ such that, if $-\e<\t<0$ and $c_0h^*\le \ell\le (c_0|\t|)^{-1/(d-1)}$, then the energy $E_T$ of any bad tile $T\in\mathcal P$ satisfies
\be E_{T}({\mathcal B}_{T})\ge \ell^d e_{\rm s}(h^*)+c_2\big[n_c(T)+|\t|^{(p-d)/(p-d-1)}\ell^d\c_{\rm hole}(T)
\big],
\label{2.5bisp}\ee
where $\c_{\rm hole}(T)$ is equal to 1 if $T$ contains a hole, and 0 otherwise.  
\end{customthml}

Its proof is a rather straightforward adaptation of the one in $d=2$, and we refer to \cite[App.~D]{GLS}, where the necessary changes were described in the case $d=3$.  

Since every portion of the boundary of a good region $G_i$ is 
adjacent to a bad tile, we have the bound $\sum_{i=1}^{\mathcal N_G} |\partial G_i|\le 2 d \ell^{d-1} {\mathcal N}_B$.
In combination with the bounds above, this leads to the following generalization of Theorem \ref{thm3}.

\begin{customthm}{3'}\label{thm3p}
For given $d\geq 2$, there exist positive constants $C_0$, $C_1$, $\e$ such that, if $J_c-\e<J<J_c$ and $C_0 h^*\le \ell\le (C_0(J_c-J))^{-1/(d-1)}$, 
then for every $\underline s\in\{\pm1\}^{\mathbb  Z^d}$ and every finite set $X\subset \mathbb Z^d$,
\bea H_X(\underline s_X|\underline \s^*)&\ge& H_X(\ss^*_X|\ss^*)
+C_1\Big(N_c+(J_c-J)^{\frac{p-d}{p-d-1}}\ell^d\mathcal N_B^{\rm hole}\Big)\nonumber\\
&+&\frac12\sum_{h\neq h^*}\sum_{i=1}^{\mathcal N_G}(e_{\rm s}(h)-e_{\rm s}(h^*))
A_h(G_i),\label{1.6p}\eea
where $N_c$, $\mathcal N_B^{\rm hole}$, and $G_i$ are, respectively, the number of corners, the
number of bad tiles containing a hole, and the good regions, associated with the infinite spin configuration $\ss=(\underline s_X,\ss^*_{X^c})$ coinciding with $\underline s_X$ on 
$X$ and with $\ss^*$ on $X^c$, defined via tiling with squares of side length $\ell$ as described above.
\end{customthm}

Theorem~\ref{thm3p} implies the analogue of Theorem~\ref{thm1}, i.e., the fact that striped configurations with stripe width $h^*$ are infinite volume ground states with trivial sectors for $J$ close to $J_c$, and also the analogue of Theorem~\ref{thm2}, stating that all infinite volume ground states that are invariant under translations by $d-1$ lattice vectors are characterized by the existence of an interface  separating the cubic lattice $\ZZZ^d$ into two components, on each of which the configuration is perfectly striped. 

\vskip.4truecm

{\bf Acknowledgments.} The research leading to these results has
received funding from the European Research Council under the European
Union's Seventh Framework Programme ERC Starting Grant CoMBoS (grant
agreement n$^o$ 239694), from the Italian PRIN National
Grant {\it Geometric and analytic theory of Hamiltonian systems in finite and infinite dimensions}, 
 and the Austrian Science Fund (FWF), project Nr. P 27533-N27.
Part of this work was completed during a stay at the Erwin Schr\"odinger Institute for Mathematical Physics in Vienna (ESI program 2015 ``Quantum many-body systems, random matrices, and disorder"), whose hospitality and financial support is gratefully acknowledged. 
We also thank Joel Lebowitz and Elliott Lieb for stimulating discussions and their constant encouragement in pursuing this project.



\begin{thebibliography}{99}

\bibitem{AWMD96} Arlett, J. P. Whitehead, A. B. MacIsaac, and K. De'Bell: {\it Phase diagram for the striped phase in the two-dimensional dipolar Ising model}, Phys. Rev. B {\bf 54}, 3394 (1996).


\bibitem{Ba} P. Ball: {\it 
The Self-Made Tapestry: Pattern Formation in Nature}, Oxford University Press, 1999.

\bibitem{BCK07} M. Biskup, L. Chayes, and S. A. Kivelson: {\it On the Absence of Ferromagnetism in Typical 2D Ferromagnets}, Commun. Math. Phys. {\bf 274}, 217--231 (2007).

\bibitem{CMST06} S. A. Cannas, M. F. Michelon, D. A. Stariolo, F. A. Tamarit: {\it Ising nematic phase in ultrathin magnetic films: A Monte Carlo study
},  Phys. Rev. B {\bf 73}, 184425 (2006).

\bibitem{CDSN12} S. Chakrabarty, V. Dobrosavljevic, A. Seidel, and
Z. Nussinov: {\it Universality of modulation length and time exponents},
Phys. Rev. E {\bf 86},  041132 (2012).

\bibitem{CN11} S. Chakrabarty and
Z. Nussinov: {\it Modulation and correlation lengths in systems with
competing interactions}, Phys. Rev. B {\bf 84}, 144402 (2011).

\bibitem{CEKNT96} L. Chayes, V. Emery, S. Kivelson, Z. Nussinov, and G. Tarjus: {\it Avoided critical behavior in a uniformly frustrated system}, Physica A {\bf 225}, 129 (1996).

\bibitem{CPPV09} F. Cinti, O. Portmann, D. Pescia, and A. Vindigni: {\it One-dimensional Ising ferromagnet frustrated by long-range interactions at finite temperatures}, Phys. Rev. B {\bf 79}, 214434 (2009).

\bibitem{CV89} R. Czech and J. Villain: {\it  Instability of two-dimensional Ising ferromagnets with dipole interactions}, J. Phys. Condens. Matter {\bf 1}, 619 (1989). 
 

\bibitem{DMW00} K. DeÕBell, A. B. MacIsaac, and J. P. Whitehead: {\it Dipolar effects in magnetic thin 
films and quasi-two-dimensional systems}, Rev. Mod. Phys. {\bf 72}, 225 (2000).

\bibitem{EJ10} E. Edlund and M. Nilsson Jacobi: {\it Universality of Striped Morphologies},
 Phys. Rev. Lett. {\bf 105}, 137203 (2010). 

\bibitem{EKT99}      V. J. Emery, S. A. Kivelson, and J. M. Tranquada: 
{\it Stripe phases in high-temperature superconductors}, Proc. Nat. Ac. Sc. USA {\bf 96}, 8814-8817
(1999).

\bibitem{FT15}     L. C. Flatley, F. Theil: {\it 
 Face-Centered Cubic Crystallization of Atomistic Configurations}, 
 Arch. Rational Mech. Anal. {\bf 218}, 363-416 (2015).

\bibitem{FK99} E. Fradkin, S. A. Kivelson: {\it Liquid-crystal phases of quantum Hall systems},
Phys. Rev. B {\bf 59}, 8065 (1999).

\bibitem{GLL06} A. Giuliani, J. Lebowitz, E. Lieb: {\it Ising models with long-range dipolar and short 
range ferromagnetic interactions}, Phys. Rev. B {\bf 74}, 064420 (2006).

\bibitem{GLL07} A. Giuliani, J. Lebowitz, E. Lieb: {\it Striped phases in two-dimensional dipole systems},
Phys. Rev. B {\bf 76}, 184426 (2007).

\bibitem{GLL09a} A. Giuliani, J. Lebowitz, E. Lieb: {\it Periodic minimizers in 1D local mean field theory}, 
Commun. Math. Phys. {\bf 286}, 163Ð177 (2009).

\bibitem{GLL09b} A. Giuliani, J. Lebowitz, E. Lieb: {\it Modulated phases of a 1D sharp interface model 
in a magnetic field}, Phys. Rev. B {\bf 80}, 134420 (2009).

\bibitem{GLL11} A. Giuliani, J. Lebowitz, E. Lieb: {\it Checkerboards, stripes and corner energies in spin models with
competing interactions}, Phys. Rev. B {\bf 84}, 064205 (2011).

\bibitem{GLS} A. Giuliani, E. H. Lieb and R. Seiringer: {\it 
Formation of Stripes and Slabs Near the Ferromagnetic Transition}, Comm. Math. Phys. {\bf 331}, 
333--350 (2014); and 
{\it Realization of stripes and slabs in two and three dimensions}, Phys. Rev. B {\bf  88}, 064401 (2013). 

\bibitem{GMu12} A. Giuliani, S. M\"uller: {\it Striped periodic minimizers of a two-dimensional model for martensitic phase
transitions}, Commun. Math. Phys. {\bf 309}, 313--339 (2012).

\bibitem{GTV00} M. Grousson, G. Tarjus, and P. Viot:{\it Phase diagram of an Ising model with long-range frustrating interactions: A theoretical analysis}, Phys. Rev. E {\bf 62}, 7781 (2000).

\bibitem{HAC00} C. Harrison et al: {\it Mechanisms of Ordering in Striped Patterns},
 Science {\bf 24}, 1558 (2000).
 
 \bibitem{HR80} R. C. Heitmann and C. Radin: {\it The ground state for sticky disks}, J. Stat. Phys. 
{\bf 22}, 281 (1980).

\bibitem{KL86} T. Kennedy and E. H. Lieb: {\it An itinerant electron model with crystalline or magnetic 
long range order}, Physica A {\bf 138}, 320 (1986).
  
\bibitem{KM} R. V. Kohn and S. M\"uller: {\it Branching of twins near an austenite--twinned-martensite 
interface}, Philos. Mag. A {\bf 66}, 697 (1992).

\bibitem{LEFK94} U. Low, V. J. Emery, K. Fabricius, and S. A. Kivelson: {\it Study of an Ising model with competing long- and short-range interactions}, Phys. Rev. Lett. {\bf 72}, 1918 (1994).

\bibitem{MWRD95} A. B. MacIsaac, J. P. Whitehead, M. C. Robinson, and K. DeÕBell: {\it Striped phases in two-dimensional dipolar ferromagnets}, Phys. Rev. B {\bf 51}, 16033 (1995).

\bibitem{MSN15} A. Mendoza-Coto, D. A. Stariolo, L. Nicolao: {\it 
Nature of long range order in stripe forming systems with long range repulsive interactions},
Phys. Rev. Lett. {\bf 114},  116101 (2015).

\bibitem{NBH08} E. Nielsen, R. N. Bhatt, and D. A. Huse:
{\it Modulated phases in magnetic models frustrated by long-range interactions}, Phys. Rev. B {\bf 77}, 054432 (2008).

\bibitem{OMYT13}  M. Okamoto, T. Maruyama, K. Yabana, and T. Tatsumi: {\it Nuclear ``pasta''
structures in low-density nuclear matter and properties of the neutron-star crust}, 
Phys. Rev. C {\bf 88}, 025801 (2013).

\bibitem{OTC09} O. Osenda, F. A. Tamarit, and S. A. Cannas:
{\it Nonequilibrium structures and slow dynamics in a two-dimensional spin system with competing long-range and short-range interactions}, , Phys. Rev. E {\bf 80}, 021114 (2009).

\bibitem{PC07} S. A. Pighin and S. A. Cannas: {\it Phase diagram of an Ising model for ultrathin magnetic films: Comparing mean field and Monte Carlo predictions}, Phys. Rev. B {\bf 75}, 224433 (2007).

\bibitem{PGSBPV10} O. Portmann, A. Golzer, N. Saratz, O. V. Billoni, D. Pescia, and A. Vindigni: {\it Scaling hypothesis for modulated systems}, Phys. Rev. B {\bf 82}, 184409 (2010).

\bibitem{RRT06} E. Rastelli, S. Regina, and A. Tassi : {\it Phase transitions in a square Ising model with exchange and dipole interactions}, Phys. Rev. B {\bf 73}, 144418 (2006).

\bibitem{SW92} M. Seul and R. Wolf: {\it Evolution of disorder in two-dimensional stripe patterns: ``Smectic'' instabilities and disclination unbinding}, Phys. Rev. Lett. {\bf 68}, 2460 (1992).

\bibitem{SK04} B. Spivak and S. A. Kivelson: {\it Phases intermediate between a two-dimensional 
electron liquid and Wigner crystal}, Phys. Rev. B {\bf 70}, 155114 (2004).

\bibitem{SS00} A. D. Stoycheva and S. J. Singer: {\it 
Stripe Melting in a Two-Dimensional System with Competing Interactions},
Phys. Rev. Lett. {\bf 84}, 4657 (2000).

\bibitem{S05} A. S\"uto: {\it Crystalline Ground States for Classical Particles}, Phys. Rev. Lett. {\bf 95}, 
265501 (2005).

\bibitem{T06} F. Theil: {\it A Proof of Crystallization in Two Dimensions}, Commun. Math. Phys. {\bf 262}, 
209 (2006).

\bibitem{VSPPP08} A. Vindigni, N. Saratz, O. Portmann, D. Pescia, and P. Politi: 
  {\it Stripe width and nonlocal domain walls in the two-dimensional dipolar frustrated Ising ferromagnet}, 
  Phys. Rev. B {\bf 77}, 092414 (2008).

\end{thebibliography}
\end{document}